\documentclass[a4paper,10pt]{article}
\usepackage[utf8x]{inputenc}
\usepackage[top=0.7in, bottom=1 in, left=1.5in, right=1.5in]{geometry}

\usepackage{graphicx}
\usepackage{tabularx}
\usepackage{booktabs}
\usepackage{pdflscape}
\usepackage{lscape}
\usepackage{rotating}
\usepackage{float}
\usepackage{makeidx}
\usepackage{colortbl}
\usepackage{longtable}
\usepackage{subcaption}
\usepackage{authblk}
\usepackage{placeins}
\usepackage{cite}
\usepackage{bbm}
\usepackage{epstopdf}
\usepackage{amsmath}
\usepackage{multirow}

\title{What does past correlation structure tell us about the future? An answer from network filtering}

\author[1]{Nicol\'o Musmeci}
\author[2,3,*]{Tomaso Aste}
\author[1,2]{T. Di Matteo}
\affil[1]{Department of Mathematics, King's College London, The Strand, London, WC2R 2LS}
\affil[2]{Department of Computer Science, UCL, Gower Street, London, WC1E 6BT, UK}
\affil[3]{Systemic Risk Centre, London School of Economics and Political Sciences, London, WC2A2AE, UK}

\affil[*]{t.aste@ucl.ac.uk}

\begin{document}

\maketitle

\begin{abstract}
We discovered that past changes in the market correlation structure are significantly related with future changes in the market volatility. By using correlation-based information filtering networks we device a new tool for forecasting the market volatility changes. 
In particular, we introduce a new measure, the ``correlation structure persistence'', that quantifies the rate of change of the market dependence structure. This measure shows a deep interplay with changes in volatility and we demonstrate it can anticipate market risk variations. 
Notably, our method overcomes the curse of dimensionality that limits the applicability of traditional econometric tools to portfolios made of a large number of assets. We report on forecasting performances and statistical significance of this tool for two different equity datasets. We also identify an optimal region of parameters in terms of True Positive and False Positive trade-off, through a ROC curve analysis. We find that our forecasting method is robust and it outperforms predictors based on past volatility only. Moreover the temporal analysis indicates that our method is able to adapt to abrupt changes in the market, such as financial crises, more rapidly than methods based on past volatility.   
\end{abstract}

\flushbottom
\maketitle
%
%
\thispagestyle{empty}


\section*{Introduction}

Forecasting changes in volatility is essential for risk management, asset pricing and scenario analysis. Indeed, models for describing and forecasting the evolution of volatility and covariance among financial assets are widely applied in industry \cite{risk_forecasting_book,multivariate_models,bouchaud_book,preis_stress}. Among the most popular approaches are worth mentioning the multivariate extensions of GARCH \cite{multigarch_survey}, the
stochastic covariance models \cite{stochastic_volatility} and realized covariance \cite{realized_volatility}.
However most of these econometrics tools are not able to cope with more than few assets, due to the curse of dimensionality and the increase in the number of parameters \cite{risk_forecasting_book}, limiting their insight into the volatility evolution to baskets of few assets only. This is unfortunate, since gathering insights into systemic risk and the unfolding of financial crises require modelling the evolution of entire markets which are composed by large numbers of assets \cite{risk_forecasting_book}. 

We suggest to use network filtering \cite{mantegna1,Tumminello05,dynamic_networks_correlation,asset_graphs,clustering_dyn_asset_graph,buccheri,caldarelli_book} as a valuable tool to overcome this limitation. 
Correlation-based filtering networks are tools which have been widely applied to filter and reduce the complexity of covariance matrices made of large numbers of assets (of the order of hundreds), representative of entire markets. This strand of research represents an important part of the Econophysics literature and has given important insights for risk management, portfolio optimization and systemic risk regulation \cite{black_monday,cluster_portfolio,bonanno_mst,invest_periph,musmeci_DBHT,musmeci_jntf}.

The volatility of a portfolio depends on the covariance matrix of the corresponding assets \cite{markowitz}.
Therefore, the latter can provide insights into the former. In this work we elaborate on this connection: we show that correlation matrices can be used to predict variations of volatility, once they are analysed through the lens of network filtering. This is quite an innovative use of correlation-based networks, which have been used mostly for descriptive analyses, with the connections with risk forecasting being mostly overlooked. Some works have shown that is possible to use dimensionality reduction techniques, such as spectral methods \cite{rmt_1}, as early-warning signals for systemic risk \cite{pca_systemic_risk,pca_systemic_risk2}: however these approaches, although promising, do not provide proper forecasting tools, as they are affected by high false positive ratios and are not designed to predict a specific quantity.

The approach we propose exploits network filtering to explicitly predict future volatility of markets made of hundreds of stocks. To this end, we introduce a new dynamical measure that quantifies the rate of change in the structure of the market correlation matrix: the ``correlation structure persistence'' $\langle ES \rangle$. This quantity is derived from the structure of network filtering from past correlations.
Then we show how such measure exhibits significant predicting power on the market volatility, providing a tool to forecast it. We assess the reliability of this forecasting through out-of-sample tests on two different equity datasets.

The rest of this paper is structured as follows: we first describe the two datasets we have analysed and we introduce the correlation structure persistence;
then we show how our analyses point out a strong interdependence between correlation structure persistence and future changes in the market volatility; moreover, we describe how this result can be exploited to provide a forecasting tool useful for risk management, by presenting out-of-sample tests and false positive analysis; then we investigate how the forecasting performance changes in time; finally we discuss our findings and their theoretical implications.

\section*{Results}

\subsection*{A measure of correlation structure persistence}
We have analysed two different datasets of equity data. The first set (NYSE dataset) is composed by daily closing prices of $N=342$ US stocks traded in New York Stock Exchange, covering 15 years from 02/01/1997 to 31/12/2012. The second set (LSE dataset) is composed by daily closing prices of $N=214$ UK stocks traded in the London Stock Exchange, covering 13 years from 05/01/2000 to 21/08/2013. All stocks have been continuously traded throughout these periods of time. These two sets of stocks have been chosen in order to provide a significant sample of the different industrial sectors in the respective markets.

For each asset $i$ ($i=1, ... ,N$) we have calculated the corresponding daily log-return $r_i(t)=log(P_i(t))-log(P_i(t-1))$, where $P_i(t)$ is the asset $i$ price at day $t$. The market return $r_M(t)$ is defined as the average of all stocks returns: $r_M(t)=1/N \sum_i r_i(t)$. In order to calculate the correlation between different assets we have then analysed the observations by using $n$ moving time windows, $T_a$ with $a=1,...,n$. Each time window contains $\theta$ observations of log-returns for each asset, totaling to $N \times n$ observations. The shift between adjacent time windows is fixed to $dT = 5$ trading days. We have calculated the correlation
matrix within each time window, $\{ \rho_{ij}(T_a) \}$, by using an exponential smoothing method \cite{exp_smoothing} that allows to assign more weight on recent observations. The smoothing factor of this scheme has been chosen equal to $\theta/3$ according to previously established criteria \cite{exp_smoothing}.

From each correlation matrix $\{ \rho_{ij}(T_a) \}$ we have then computed the corresponding Planar Maximally Filtered Graph (PMFG) \cite{PMFG2}. The PMFG is a sparse network representation of the correlation matrix that retains only a subset of most significant entries, selected through the topological criterion of being maximally planar \cite{Tumminello05}. Such networks serve as filtering method and have been shown to provide a deep insight into the dependence structure of financial assets \cite{Tumminello05,dynamic_networks_correlation,NJP10}. 

\begin{figure}[t!]
    \includegraphics[trim = 5mm 10mm 20mm 65mm, clip, scale=0.6]{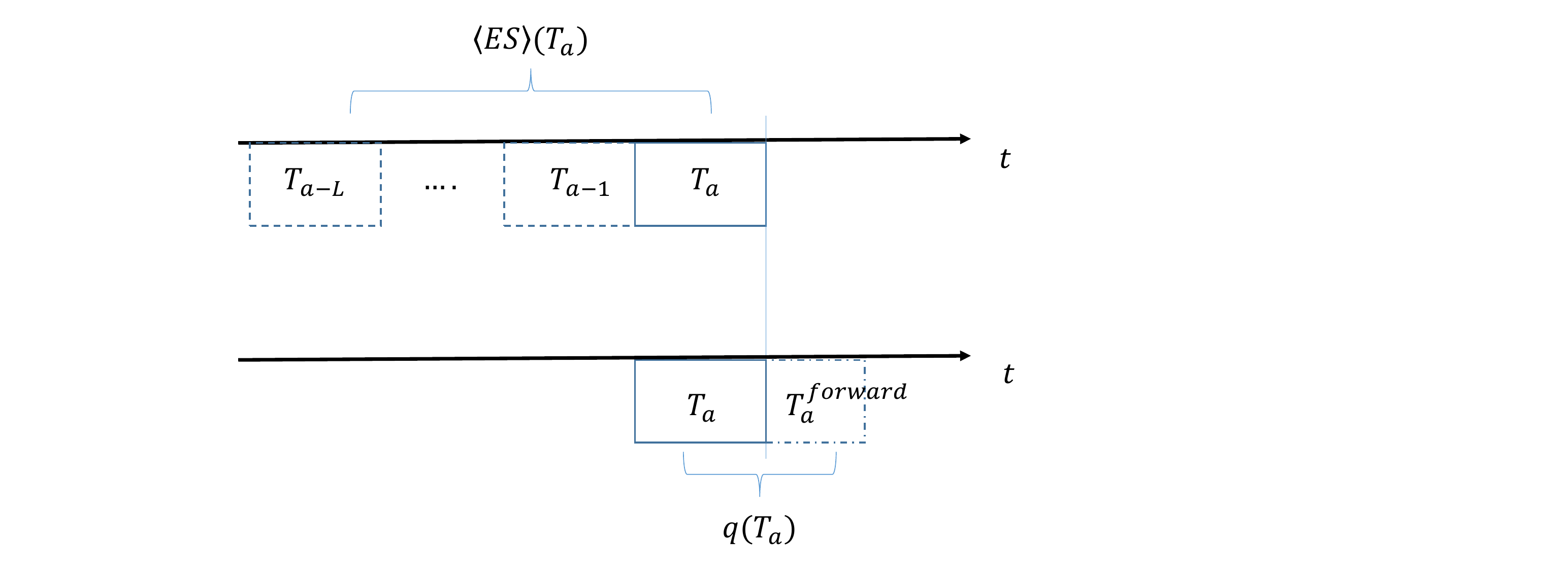}
\caption{\label{fig:time_windows} {\bf Scheme of time windows setting for $\langle ES \rangle (T_a)$ and $q(T_a)$ calculation}. $T_a$ is a window of length $\theta$. The correlation structure persistence $\langle ES \rangle(T_a)$ (upper axis) is computed by using data in $T_a$ and in the first $L$ time windows before $T_a$. The volatility ratio $q(T_a)$ is computed by using data in $T_a$ and in the future time window $T_a^{forward}$. In the upper axis the time windows are actually overlapping, but they are here represented as disjoint for the sake of simplicity.}
\end{figure}

Once the $n$ PMFGs, $G(T_a)$ with $a=1,...,n$, have been computed we have calculated two measures, a backward-looking and a forward-looking one. The first is a measure that monitors the
correlation structure persistence, based on a measure of PMFG similarity. 
This backward-looking measure, that we call $\langle ES \rangle (T_a)$, relies on past data only and indicates how slowly the correlation structure measured at time window $T_a$ is differentiating from structures associated to previous time windows. The forward-looking measure is the volatility ratio $q(T_a)$ \cite{kondor,kondor2}, that at each time window quantifies how good the market volatility measured at $T_a$ is as a proxy for the next time-window volatility. Unlike $\langle ES \rangle (T_a)$, the value of $q(T_a)$ is not known at the end of $T_a$. Fig. \ref{fig:time_windows} shows a graphical representation of the time window set-up. In the following we define the two measures:

\label{sec:methods}

\begin{itemize}
 \item {\bf Correlation structure persistence $\langle ES \rangle (T_a)$}: we define the correlation structure persistence at time $T_a$ as:

\begin{equation}
 \langle ES \rangle (T_a) = \sum_{b = a -L }^{a - 1} \omega(T_b) ES(T_a, T_b) ,
 \label{eq:es}
\end{equation}
where $\omega(T_b) = \omega_0 \exp(\frac{b-a-1}{L/3})$ is an exponential smoothing factor, $L$ is a parameter and $ES(T_a, T_b)$ is the fraction of edges in common between the two PMFGs $G(T_a)$ and $G(T_b)$, called ``edge survival ratio'' \cite{black_monday}. In formula, $ES(T_a, T_b)$ reads: 

\begin{equation}\label{eq:es_ab}
 ES(T_a, T_b) = \frac{1}{N_{edges}} \mid E^{T_a} \cap E^{T_b} \mid ,
\end{equation}
where $N_{edges}$ is the number of edges (links) in the two PMFGs (constant and equal to $3N-6$ for a PMFG \cite{PMFG2}), and $E^{T_{a}}$ ($E^{T_{b}}$) represents the edge-sets of PMFG at $T_{a}$ ($T_{b}$). The correlation structure persistence $\langle ES \rangle (T_a)$ is therefore a weighted average of the similarity (as measured by the edge survival ratio) between $G(T_a)$ and the first $L$ previous PMFGs, with an exponential smoothing scheme that gives more weight to those PMFGs that are closer to $T_a$. The parameter $\omega_0$ in Eq. \ref{eq:es} can be calculated by imposing $\sum_{b = a -L }^{a - 1} \omega(T_b) = 1$. Intuitively, $\langle ES \rangle (T_a)$ measures how slowly
the change of correlation structure is occurring in the near past of $T_a$.

\item {\bf Volatility ratio $q(T_a)$} \cite{kondor}: 
 In order to quantify the agreement between the estimated and the realized risk we here make use of the volatility ratio, a measure which has been used \cite{kondor,non_stationary_corr} for this purpose and defined as follows: 

\begin{equation}
\label{eq:q}
 q(T_a) = \frac{\sigma(T^{forward}_a)}{\sigma(T_a)} ,
\end{equation}
where $\sigma(T^{forward}_a)$ is the realized volatility of the average market return $r_M(t)$ computed on the time window $T^{forward}_a$; $\sigma(T_a)$ is the estimated volatility of $r_M(t)$ computed on time window $T_a$, by using the same exponential smoothing scheme \cite{exp_smoothing} described for the correlation $\{ \rho_{ij}(T_a) \}$.
Specifically, $T^{forward}_a$ is the time window of length $\theta_{forward}$ that follows immediately $T_a$: if $t_{\theta}$ is the last observation in $T_a$, $T^{forward}_a$ covers observations from $t_{\theta+1}$ to $t_{\theta+1+\theta_{forward}}$ (Fig. \ref{fig:time_windows}). Therefore the ratio in Eq. \ref{eq:q} estimates the agreement between the market volatility estimated with observations in $T_a$ and the actual market volatility observed over an investment in the $N$ assets over $T^{forward}_a$. If $q(T_a)>1$, then the historical data gathered at $T_a$ has underestimated the (future) realized volatilty, whereas $q(T_a)<1$ indicates overestimation. 

Let us stress that $q(T_a)$ provides an information on the reliability of the covariance estimation too, given the relation between market return volatility and covariance \cite{markowitz}:

\begin{equation}
\sigma(T_a) = \sqrt{\frac{1}{N^2} \sum_{ij} Cov_{ij}(T_a)} ,
\end{equation}

\begin{equation}
\sigma(T^{forward}_a) = \sqrt{\frac{1}{N^2} \sum_{ij} Cov_{ij}(T^{forward}_a)} ,
\end{equation}
where $Cov_{ij}(T_a)$ and $Cov_{ij}(T^{forward}_a)$ are respectively the estimated and realized covariances. 

%

\end{itemize}

\subsection*{Interplay between correlation structure persistence and volatility ratio}
To investigate the relation between $\langle ES \rangle (T_a)$ and $q(T_a)$ we have calculated the two quantities with different values of $\theta$ and $L$ in Eqs. \ref{eq:es} and \ref{eq:q}, to assess the robustness against these parameters. Specifically, we have used $\theta \in (250, 500, 750, 1000)$ trading days, that correspond to time windows of length 1, 2, 3 and 4 years respectively; $L \in (10, 25, 50, 100)$, that correspond (given $dT = 5$ trading days) to an average in Eq. \ref{eq:es} reaching back to $50$, $125$, $250$ and $500$ trading days respectively. $\theta_{forward}$ has been chosen equal to $250$ trading days (one year) for all the analysis. 

\begin{figure}[t!]
    \includegraphics[scale=0.4]{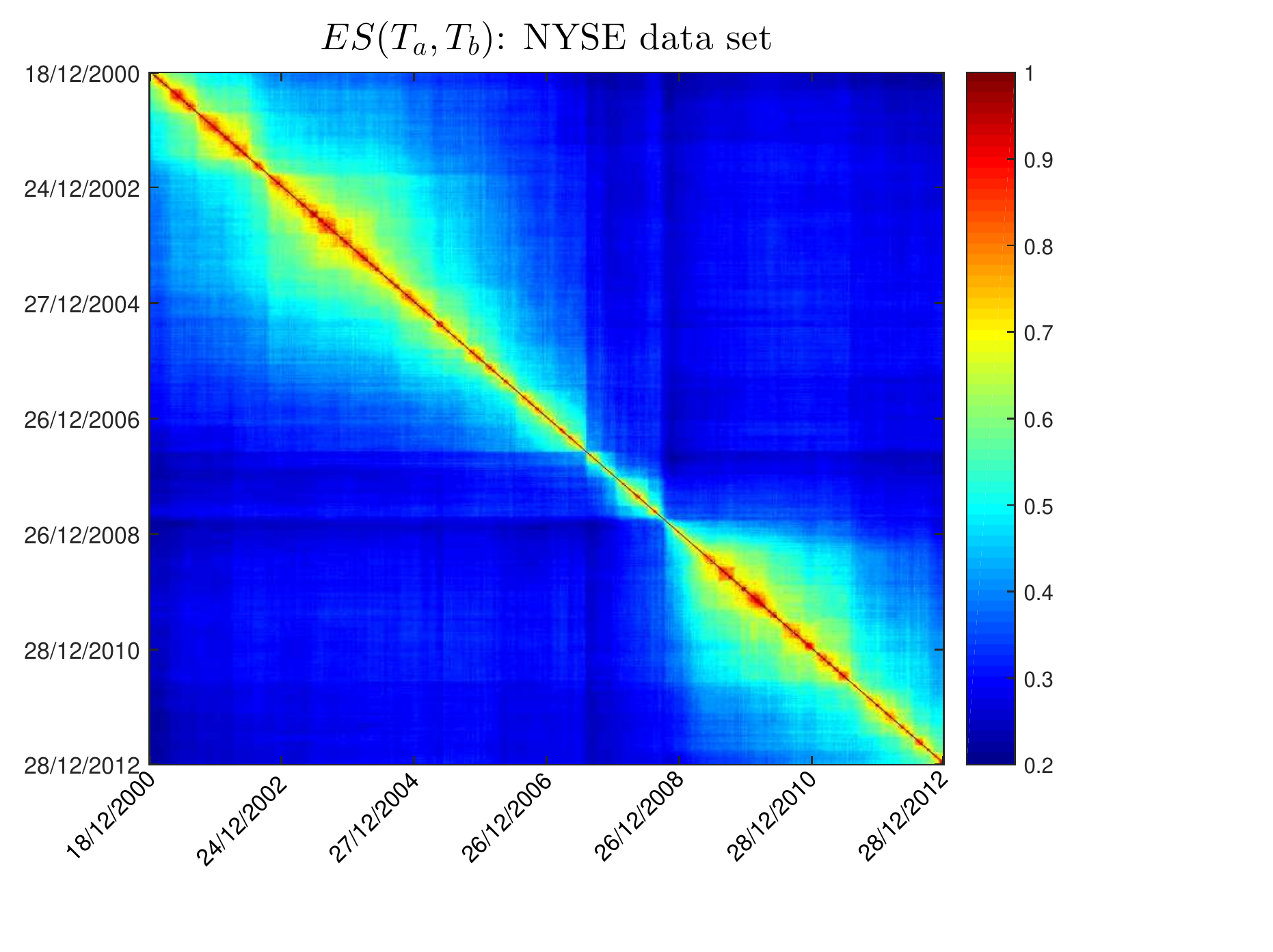}
		\hspace{-3em}
    \includegraphics[scale=0.4]{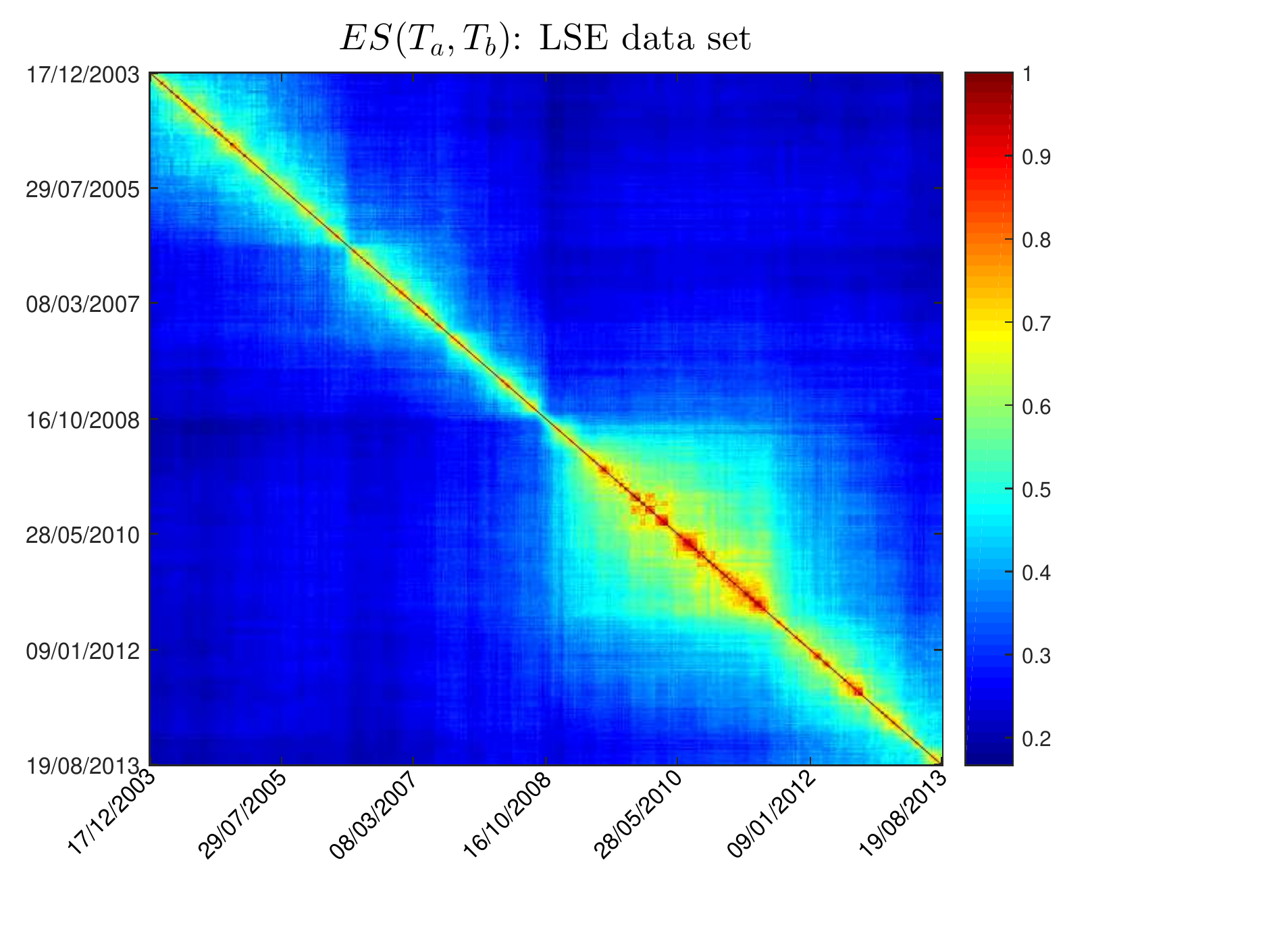}
\caption{\label{fig:es_matrices} {\bf $ES(T_a, T_b)$ matrices for $\theta = 1000$, for NYSE (left) and LSE dataset (right)}. A block-like structure can be observed in both datasets, with periods of high structural persistence and other periods whose correlation structure is changing faster. The 2007-2008 financial crisis marks a transition between two main blocks of high structural persistence.}
\end{figure}

\begin{figure}[t!]
    \includegraphics[scale=0.5]{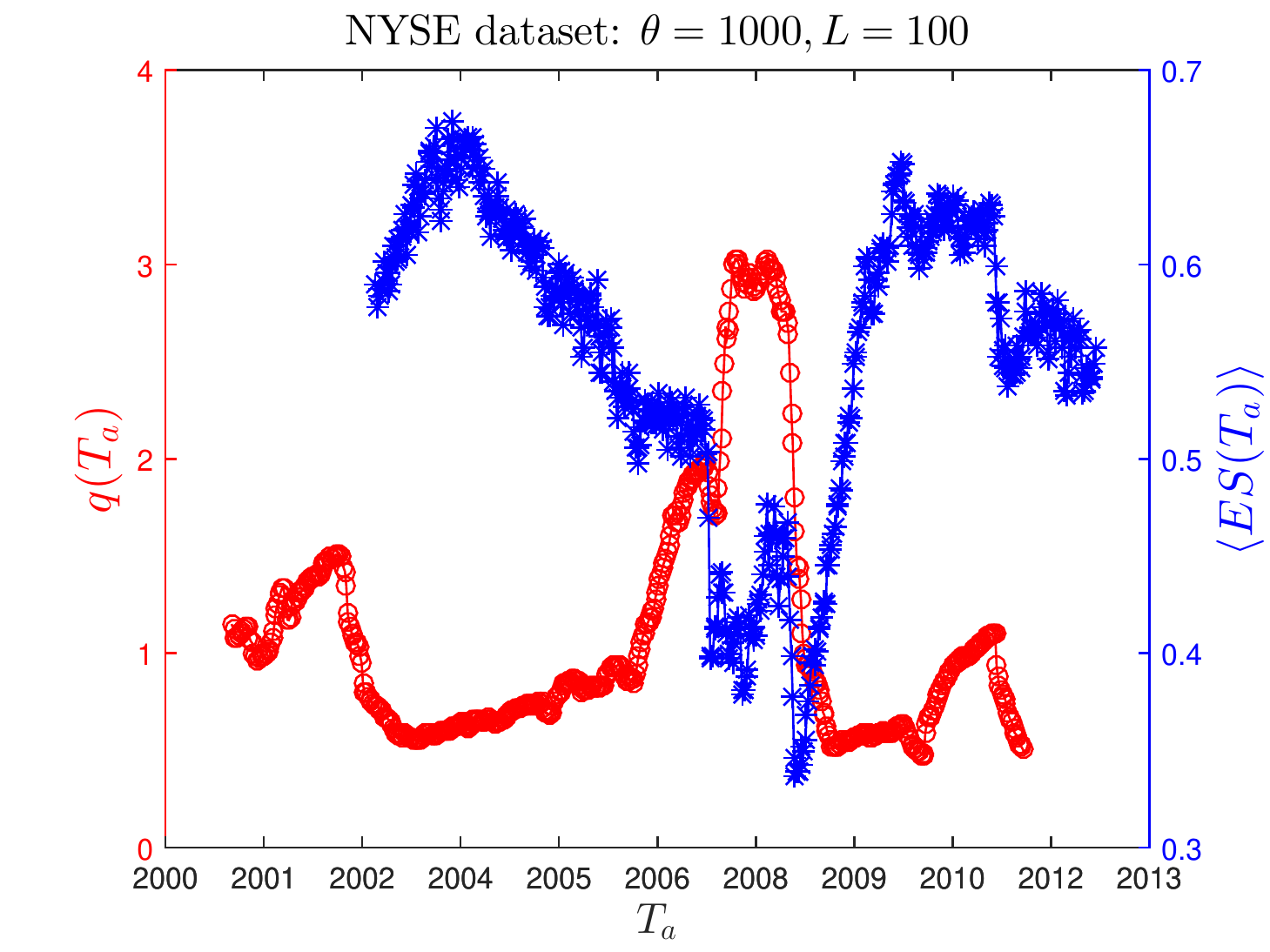}
    \includegraphics[scale=0.5]{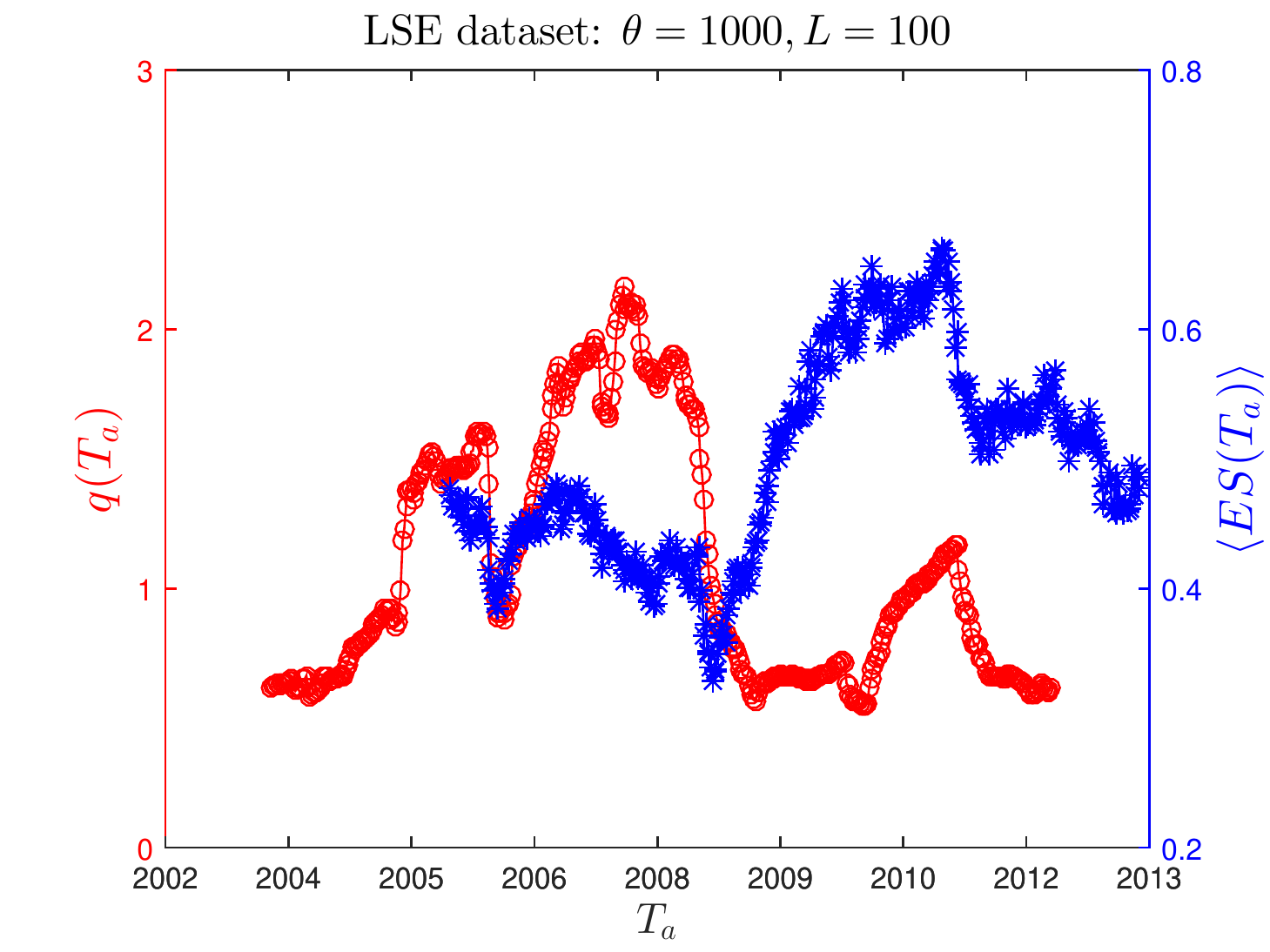}
\caption{\label{fig:plot_342} {\bf $\langle ES \rangle (T_a)$ and $q(T_a)$ signals represented for $\theta=1000$ and $L=100$}, for both NYSE (left graph) and LSE (right graph) datasets. It is evident the anticorrelation between the two signals. The financial crisis triggers a major drop in the structural persistence and a corresponding peak in $q(T_a)$.
}
\end{figure}

In Fig. \ref{fig:es_matrices} we show the $ES(T_a,T_b)$ matrices (Eq. \ref{eq:es_ab})
for the NYSE and LSE dataset, for $\theta=1000$. We can observe a block structure with periods of high structural persistence and other periods whose correlation structure is changing faster. In particular two main blocks of high persistence can be found before and after the 2007-2008 financial crisis; a similar result was found in a previous work \cite{musmeci_jntf} with a different measure of similarity. These results are confirmed for all values of $\theta$ considered.
In Fig. \ref{fig:plot_342} we show $\langle ES \rangle (T_a)$ and $q(T_a)$ as a function of time, for $\theta=1000$ and $L=100$.
As expected, main peaks of $q(T_a)$ occur during the months before the most turbulent periods in the stock market, namely the 2002 market downturn
and the 2007-08 credit crisis. Interestingly, the corresponding $\langle ES \rangle (T_a)$ seems to follow a specular trend. This is confirmed by explicit calculation of Pearson correlation between the two signals, reported in Tabs. \ref{tab:corr_4} - \ref{tab:corr_8}: as one can see, for all combinations of parameters the correlation is negative. 

 \begin{figure}[t!]
    \includegraphics[scale=0.5]{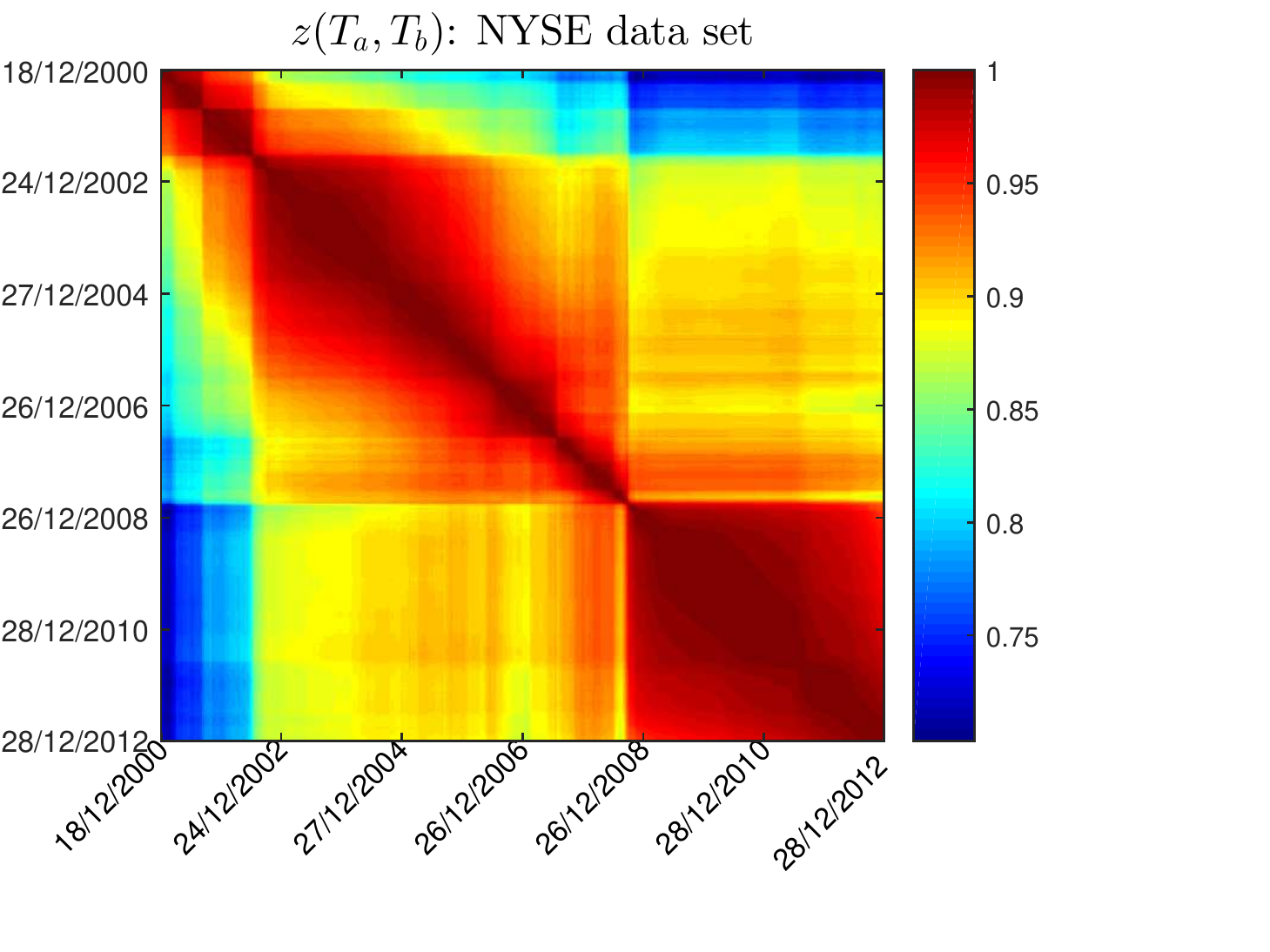}
    \includegraphics[scale=0.5]{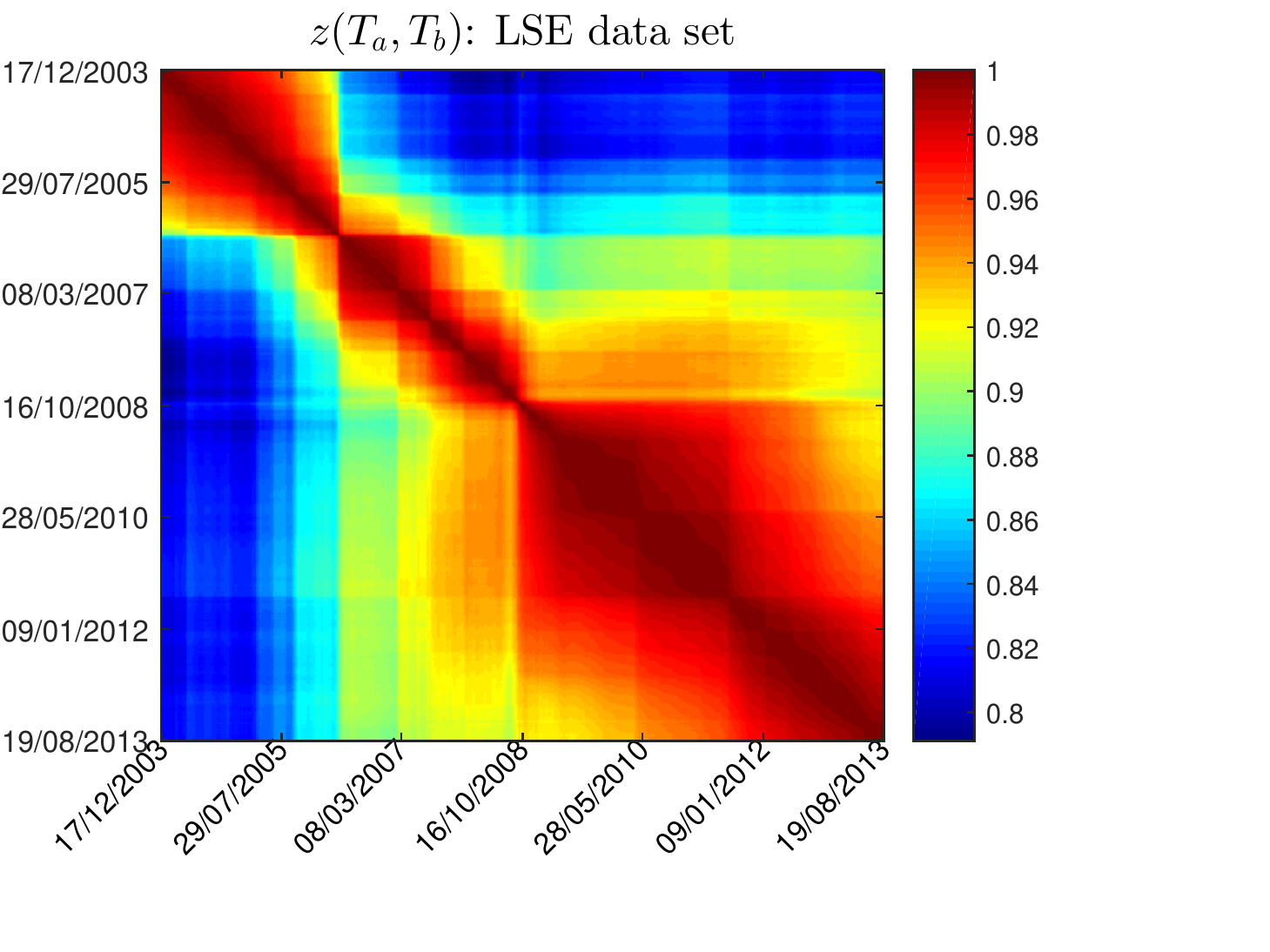}
\caption{\label{fig:meta_matrices} {\bf $z(T_a, T_b)$ matrices for $\theta = 1000$, for NYSE (left) and LSE dataset (right)}. A block-like structure can be observed in both datasets, with periods of high structural persistence and other periods whose correlation structure is changing faster. The blocks of high similarity show higher compactness than in Fig. \ref{fig:es_matrices}.}
\end{figure}

In order to check the significance of this anticorrelation we cannot rely on standard tests on Pearson coefficient, such as Fisher transform \cite{fisher_transform}, as they assume i.i.d. series \cite{math_statistics}. Our time series are instead strongly autocorrelated, due to the overlapping between adjacent time windows. Therefore we have calculated confidence intervals by performing a block bootstrapping test \cite{block_bootstrapping}. This is a variation of the bootstrapping test \cite{bootstrapping}, conceived to take into account the autocorrelation structure of the bootstrapped series. The only free parameter in this method is the block length, that we have chosen applying the optimal selection criterion proposed in literature \cite{optimal_block_bootstrapping}: such criterion is adaptive on the autocorrelation strength of the series as measured by the correlogram. We have found, depending on the parameters $\theta$ and $L$, optimal block lengths ranging from 29 to 37, with a mean of 34 (corresponding to 170 trading days). By performing block bootstrapping tests we have therefore estimated confidence intervals for the true correlation between $\langle ES \rangle (T_a)$ and $q(T_a)$; in Tabs. \ref{tab:corr_4} - \ref{tab:corr_8} correlations whose $95\%$ and $99\%$ confidence intervals (CI) do not include zero are marked with one and two stars respectively. As we can see, 14 out of 16 correlation coefficients are significantly different from zero within $95\%$ CI in the NYSE dataset, and 12 out of 16 in the LSE dataset. For what concerns the $99\%$ CI, we observe 13 out of 16 for the NYSE and 9 out of 16 for the LSE dataset. Non-significant correlations appear only for $\theta = 250$, suggesting that this length is too small to provide a reliable measure of structural persistence. Very similar results are obtained by using Minimum Spanning Tree (MST) \cite{mst_history} instead of PMFG as network filtering.        

Given the interpretation of $\langle ES \rangle (T_a)$ and $q(T_a)$ given above, anticorrelation implies that an increase in the ``speed'' of correlation structure evolution (low $\langle ES \rangle (T_a)$) is likely to correspond to underestimation of future market volatility from historical data (high $q(T_a$)), whereas when the structure evolution ``slows down'' (high $\langle ES \rangle (T_a)$) there is indication that historical data is likely to provide an overestimation of future volatility. This means that we can use
$\langle ES \rangle (T_a)$ as a valuable predictor of current historical data reliability. This result is to some extent surprising as $\langle ES \rangle (T_a)$ is derived from PMFGs topology, which in turns depends only on the ranking of correlations and not on their actual value: yet, this information provides meaningful information about the future market volatility and therefore about the future covariance.

In principle other measures of correlation ranking structure, more straightforward than the correlation persistence $\langle ES \rangle (T_a)$, might capture the same interplay with $q(T_a)$. We have therefore considered also the Metacorrelation $z(T_a,T_b)$, that is the Pearson correlation computed between the coefficients of correlation matrices at $T_a$ and $T_b$ (see Methods for more details). Such measure does not make use of PMFG. Fig. \ref{fig:meta_matrices} displays the similarity matrices obtained with this measure for NYSE and LSE datasets: we can observe again block-like structures, that however carry different information from the $ES(T_a,T_b)$ in Fig. \ref{fig:es_matrices}; in particular, blocks show higher intra-similarity and less structure. Similarly to Eq. \ref{eq:es}, we have then defined $z(T_a)$ as the weighted average over $L$ past time windows (see Methods).
%
%
%
In Tabs. \ref{tab:corr_meta_nyse} and \ref{tab:corr_meta_lse} we show the correlation between $z(T_a)$ and $q(T_a)$. As we can see, although an anticorrelation is present for each combination of parameters $\theta$ and $L$, correlation coefficients are systematically closer to zero than in Tabs. \ref{tab:corr_4} - \ref{tab:corr_8}, where $\langle ES \rangle (T_a)$ was used. Moreover the number of significant Pearson coefficients, according to the block bootstrapping, decreases to 12 out of 16 in NYSE and to 10 out of 16 in LSE dataset. Since $\langle z \rangle (T_a)$ does not make use of PMFG, this result suggests that the filtering procedure associated to correlation-based networks is a necessary step for capturing at best the correlation ranking evolution and its interplay with the volatility ratio.  


\begin{table}[h]\centering
\caption{\label{tab:corr_4} {\bf NYSE dataset: correlation between $\langle ES \rangle (T_a)$ and $q(T_a)$}, for different combinations of parameters $\theta$ and $L$. Stars mark those correlation coefficients whose confidence interval excludes zero with a $95\%$ (one star) or a $99\%$ confidence (two stars). The confidence intervals are computed from the block-bootstrapped sample.}
\begin{tabular}{ cc | c c c c |}
\cline{3-6}
& & \multicolumn{4}{c|}{L} \\
\cline{3-6}
 & & \textbf{10} & \textbf{25} & \textbf{50} & \textbf{100}\\
\hline 
\multicolumn{1}{|c}{\multirow{4}{*}{$\theta$}} & \multicolumn{1}{|c|}{\textbf{250}}  &  -0.2129  & -0.2224 &  $-0.2997^{*}$ &  $-0.3498^{**}$\\
 \multicolumn{1}{ |c  }{}   & \multicolumn{1}{|c|}{\textbf{500}}  &  $-0.4276^{**}$  & $-0.4683^{**}$  & $-0.4945^{**}$ & $-0.5354^{**}$ \\
 \multicolumn{1}{ |c  }{} & \multicolumn{1}{|c|}{\textbf{750}}  & $-0.4994^{**}$  & $-0.5499^{**}$ &  $-0.5837^{**}$ & $-0.6018^{**}$\\
 \multicolumn{1}{ |c  }{} & \multicolumn{1}{|c|}{\textbf{1000}} & $-0.5789^{**}$  & $-0.6152^{**}$  & $-0.6480^{**}$ &  $-0.6874^{**}$ \\
\hline
\addlinespace[1ex]
\multicolumn{6}{l}{\textsuperscript{**}$p<0.001$, \textsuperscript{*}$p<0.01$,}
\end{tabular}
\end{table}

\begin{table}[h]\centering
\caption{\label{tab:corr_8} {\bf LSE dataset: correlation between $\langle ES \rangle (T_a)$ and $q(T_a)$}, for different combinations of parameters $\theta$ and $L$. Stars mark those correlation coefficients whose confidence interval excludes zero with a $95\%$ (one star) or a $99\%$ confidence (two stars). The confidence intervals are computed from the block-bootstrapped sample.}
\begin{tabular}{ cc | c c c c |}
\cline{3-6}
& & \multicolumn{4}{c|}{L} \\
\cline{3-6}
 & & \textbf{10} & \textbf{25} & \textbf{50} & \textbf{100}\\
\hline 
\multicolumn{1}{|c}{\multirow{4}{*}{$\theta$}} & \multicolumn{1}{|c|}{\textbf{250}}  &  $-0.2084^{*}$  & $-0.1887^{*}$  & -0.1872 &  $-0.2269^{*}$\\
 \multicolumn{1}{ |c  }{}   & \multicolumn{1}{|c|}{\textbf{500}}  &  $-0.3083^{**}$ &  $-0.3343^{**}$  & $-0.3782^{**}$ &  $-0.4202^{**}$\\
 \multicolumn{1}{ |c  }{} & \multicolumn{1}{|c|}{\textbf{750}}  & $-0.4050^{**}$  & $-0.4409^{**}$ &  $-0.4334^{**}$  & $-0.4374^{**}$\\
 \multicolumn{1}{ |c  }{} & \multicolumn{1}{|c|}{\textbf{1000}} &  $-0.4552^{**}$ &  $-0.5285^{**}$ &  $-0.5480^{**}$  & $-0.5227^{**}$\\
\hline
\addlinespace[1ex]
\multicolumn{6}{l}{\textsuperscript{**}$p<0.001$, \textsuperscript{*}$p<0.01$,}
\end{tabular}
\end{table}

\begin{table}[h]\centering
\caption{\label{tab:corr_meta_nyse} {\bf NYSE dataset: correlation between $\langle z \rangle (T_a)$ and $q(T_a)$}, for different combinations of parameters $\theta$ and $L$. Stars mark those correlation coefficients whose confidence interval excludes zero with a $95\%$ (one star) or a $99\%$ confidence (two stars). The confidence intervals are computed from the block-bootstrapped sample.}
\begin{tabular}{ cc | c c c c |}
\cline{3-6}
& & \multicolumn{4}{c|}{L} \\
\cline{3-6}
 & & \textbf{10} & \textbf{25} & \textbf{50} & \textbf{100}\\
\hline 
\multicolumn{1}{|c}{\multirow{4}{*}{$\theta$}} & \multicolumn{1}{|c|}{\textbf{250}}  &  -0.0992   & -0.0754   & -0.1055  & -0.1157\\
 \multicolumn{1}{ |c  }{}   & \multicolumn{1}{|c|}{\textbf{500}}  & -0.2146  &  -0.2232 & -0.2309 & -0.2753\\
 \multicolumn{1}{ |c  }{} & \multicolumn{1}{|c|}{\textbf{750}}  &  -0.2997  & $-0.3706^{*}$  & $-0.4030^{*}$ & $-0.4109^{*}$\\
 \multicolumn{1}{ |c  }{} & \multicolumn{1}{|c|}{\textbf{1000}} &  $-0.3933^{**}$  & $-0.4290^{**}$  & $-0.4678^{**}$ & $-0.4574^{*}$\\
\hline
\addlinespace[1ex]
\multicolumn{6}{l}{\textsuperscript{**}$p<0.001$, \textsuperscript{*}$p<0.01$,}
\end{tabular}
\end{table}

\begin{table}[h!]\centering
\caption{\label{tab:corr_meta_lse} {\bf LSE dataset: correlation between $\langle z \rangle (T_a)$ and $q(T_a)$}, for different combinations of parameters $\theta$ and $L$. Stars mark those correlation coefficients whose confidence interval excludes zero with a $95\%$ (one star) or a $99\%$ confidence (two stars). The confidence intervals are computed from the block-bootstrapped sample.}
\begin{tabular}{ cc | c c c c |}
\cline{3-6}
& & \multicolumn{4}{c|}{L} \\
\cline{3-6}
 & & \textbf{10} & \textbf{25} & \textbf{50} & \textbf{100}\\
\hline 
\multicolumn{1}{|c}{\multirow{4}{*}{$\theta$}} & \multicolumn{1}{|c|}{\textbf{250}}  &  -0.1470  & -0.1095  & -0.1326 & -0.1720\\
 \multicolumn{1}{ |c  }{}   & \multicolumn{1}{|c|}{\textbf{500}}  & $-0.2365^{*}$ & -0.2113  & $-0.2936^{*}$ & $-0.3932^{**}$\\
 \multicolumn{1}{ |c  }{} & \multicolumn{1}{|c|}{\textbf{750}}  &  $-0.3123^{**}$ & $-0.3379^{*}$ & $-0.3538^{*}$  & $-0.3851^{*}$\\
 \multicolumn{1}{ |c  }{} & \multicolumn{1}{|c|}{\textbf{1000}} &  $-0.2917^{*}$ & -0.2954 & -0.3163 & $-0.4192^{**}$\\
\hline
\addlinespace[1ex]
\multicolumn{6}{l}{\textsuperscript{**}$p<0.001$, \textsuperscript{*}$p<0.01$,}
\end{tabular}
\end{table} 

\subsection*{Forecasting volatility: a new approach}
\label{sec:forecasting_outS}

\begin{figure}[ht!]
    \includegraphics[scale=0.4]{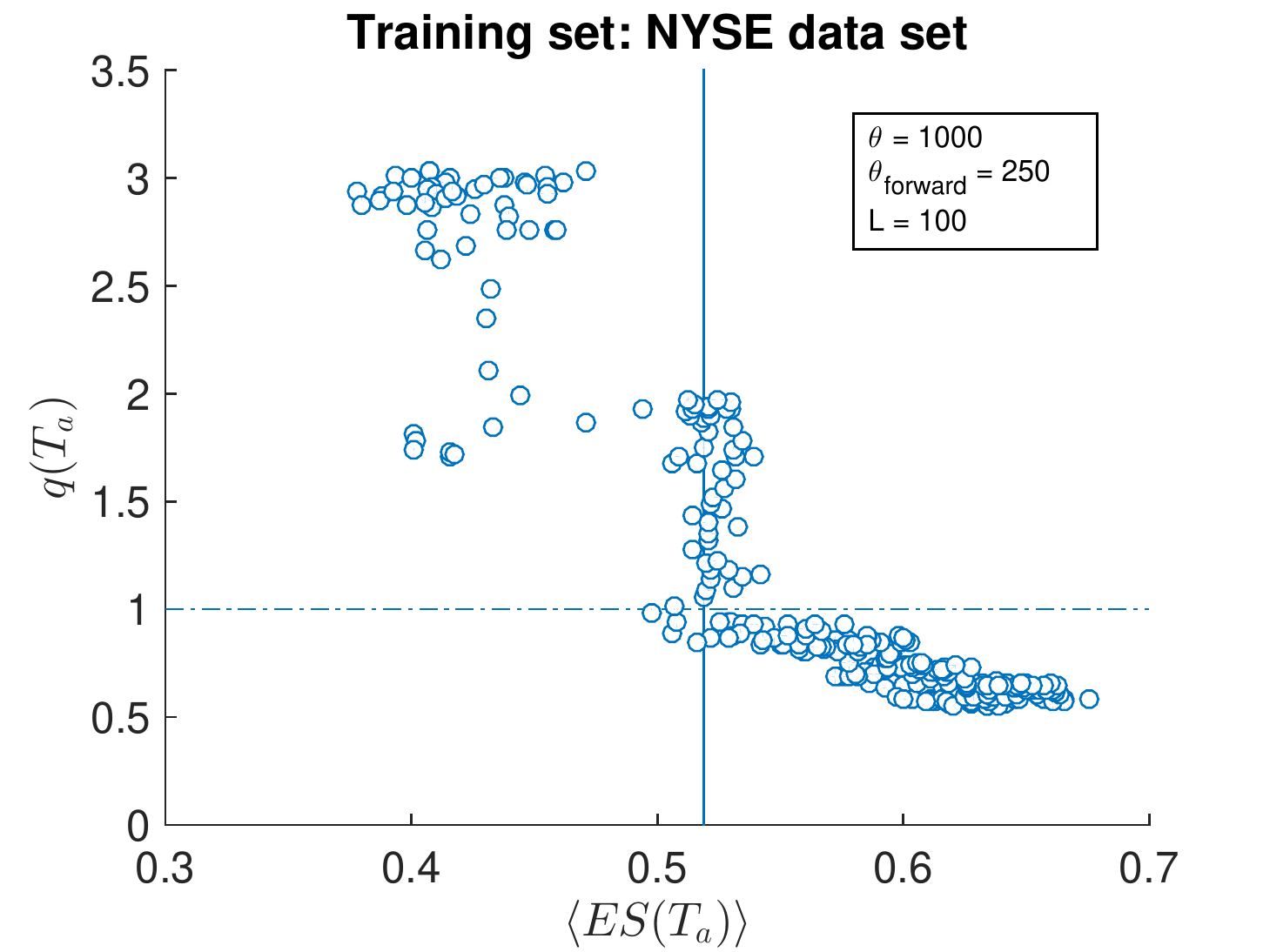}
    \includegraphics[scale=0.4]{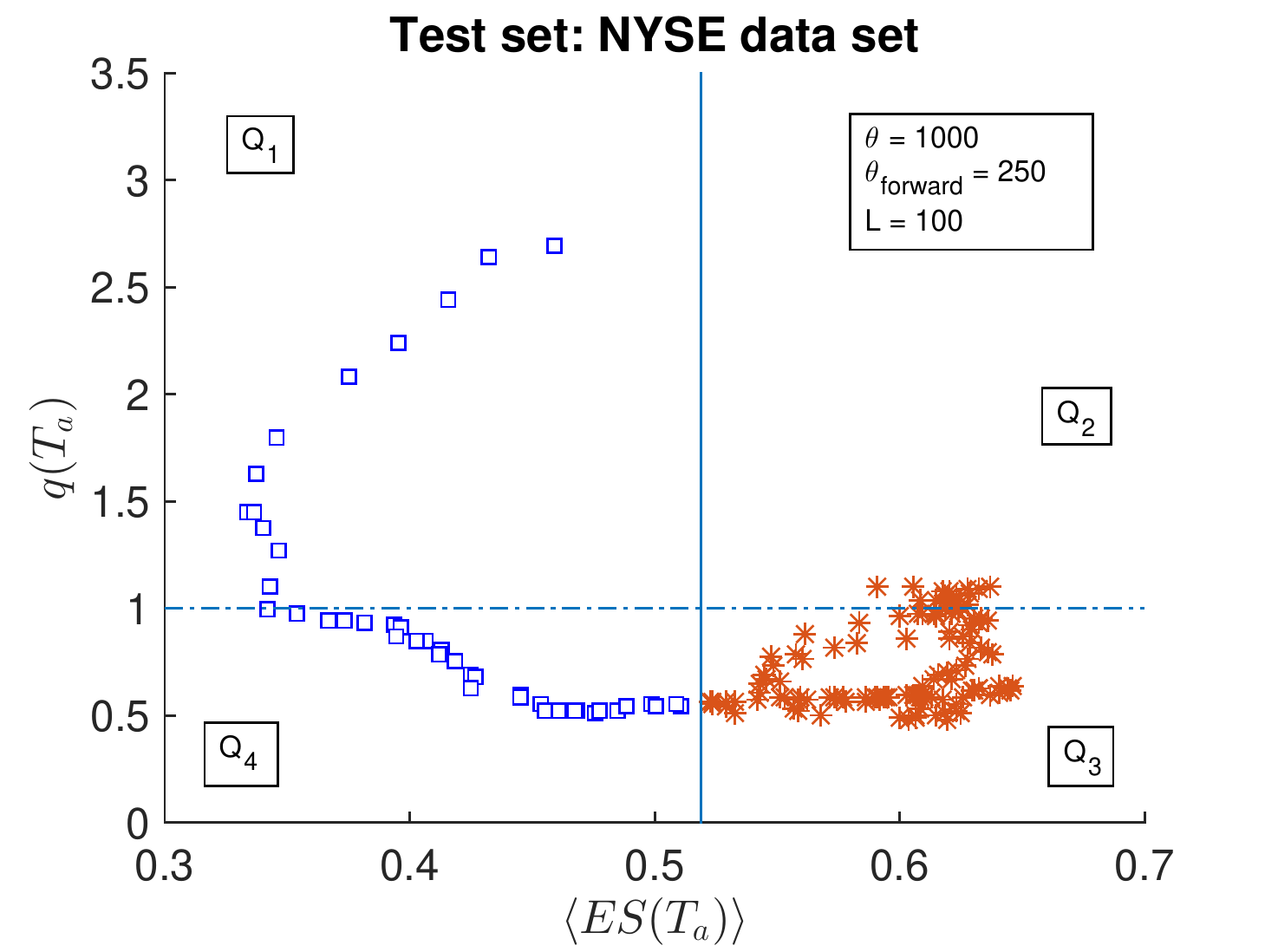}
		
    \includegraphics[scale=0.4]{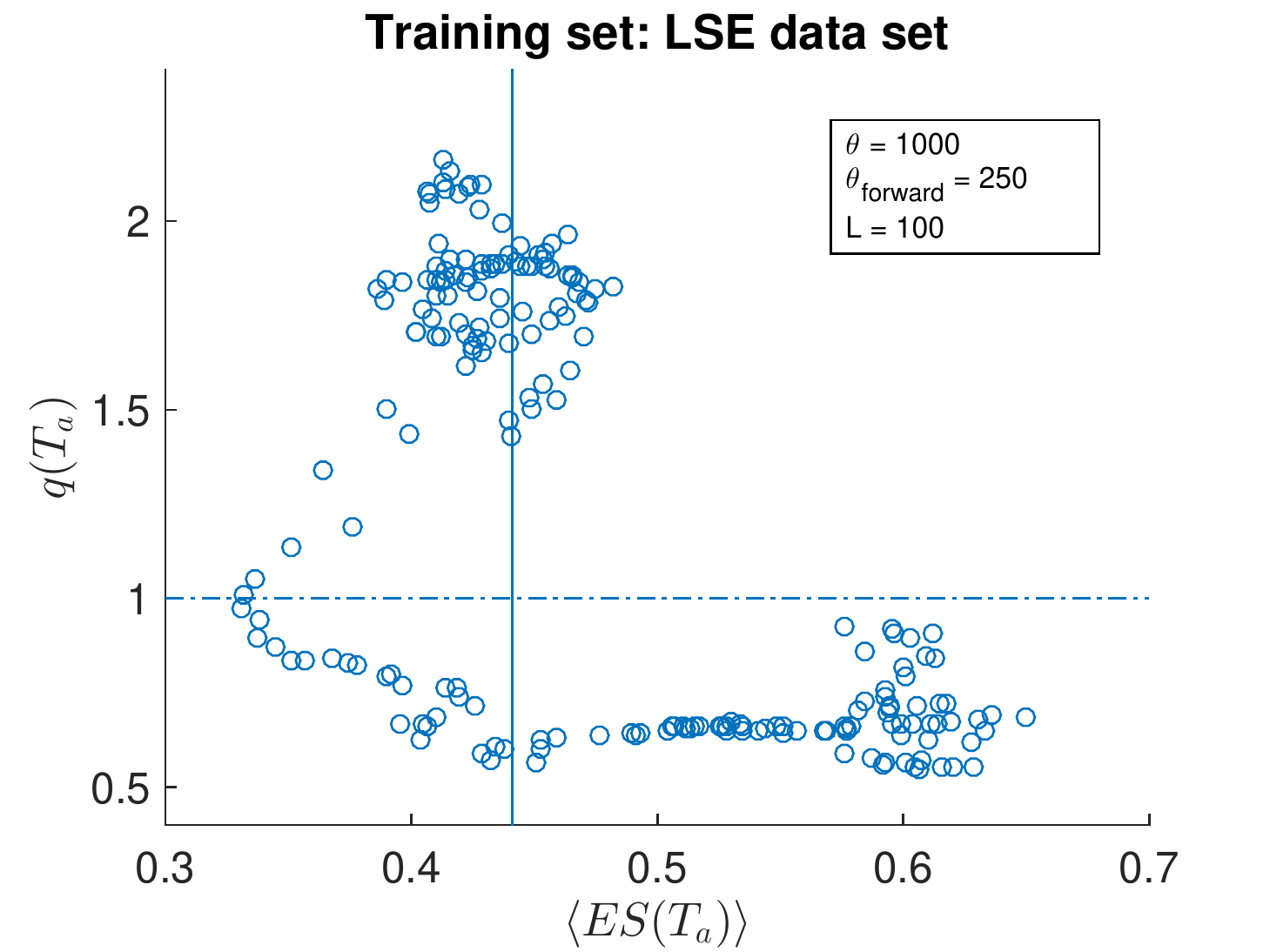}
    \includegraphics[scale=0.4]{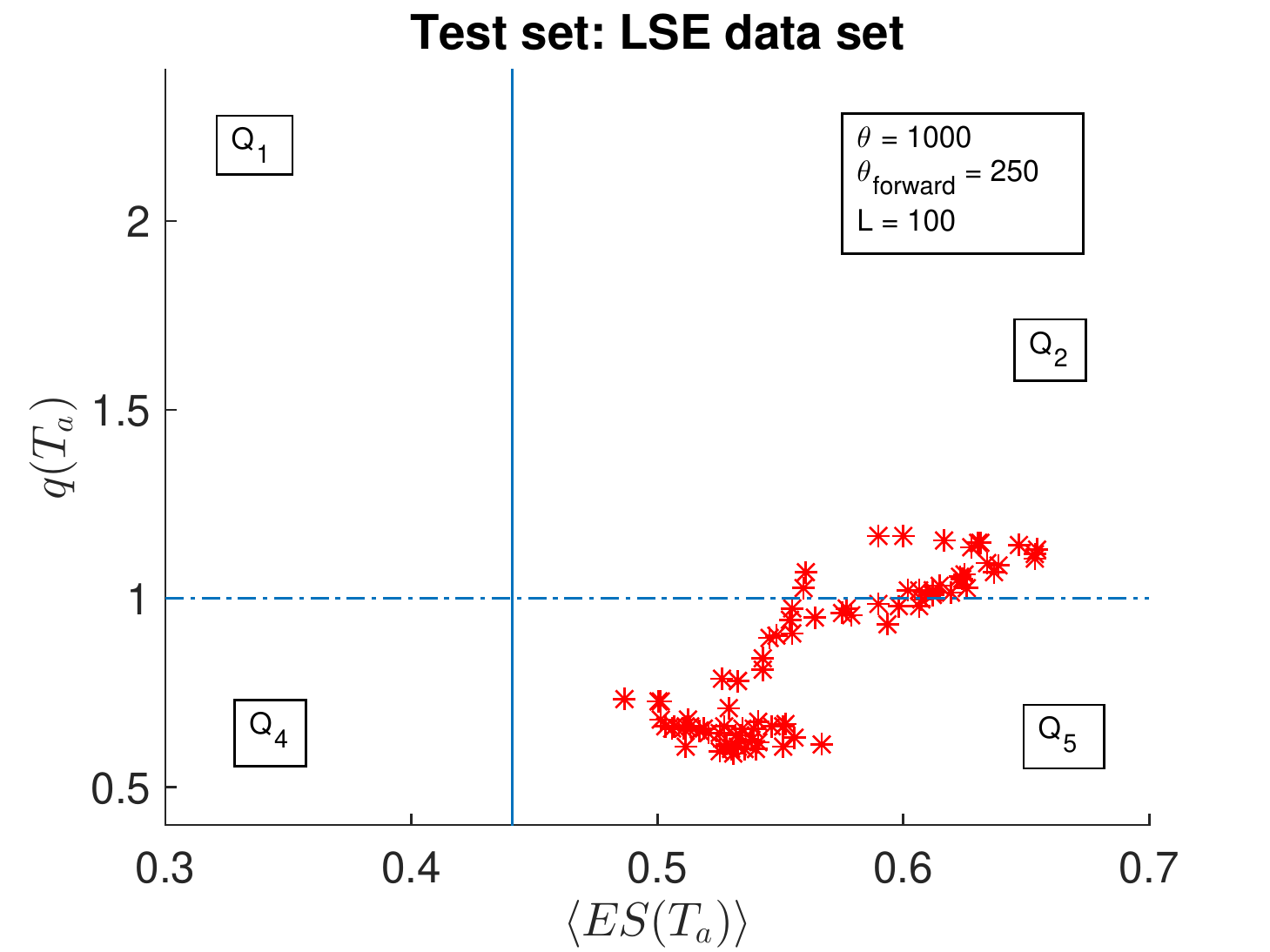}
\caption{ \label{fig:training_test} {\bf Partition of data into training (left graphs) and test (right graphs) set}. Training sets are used to regress $Y(T_a)$ against 
$\langle ES \rangle (T_a)$, in order to estimate the coefficents in the logistic regression and therefore identify the regression threshold, shown as a vertical continuous line. The test sets are used to test the forecasting performance of such regression on a subset of data that has not been used for regression; the model predicts $Y(T_a)=1$ ($q(T_a)>1$) if $\langle ES \rangle (T_a)$ is greater than the regression threshold, and $Y(T_a)=0$ ($q(T_a)<1$) otherwise.}
\end{figure}

In this section we evaluate how well the correlation structure persistence $\langle ES \rangle (T_a)$ can forecast the future through its relation with the forward-looking volatility ratio $q(T_a)$. In particular we focus on estimating whether $q(T_a)$ is greater or less than $1$: this information, although less complete than a
 precise estimation of $q(T_a)$, gives us an important insight into possible overestimation ($q(T_a)<1$) or underestimation ($q(T_a)>1$) of future volatility.

 We have proceeded as follows. Given a choice of parameters $\theta$ and $L$, we have calculated the corresponding set of pairs $\{ \langle ES \rangle (T_a) , q(T_a) \}$, with $a=1,...,n$.
 Then we have defined $Y(T_a)$ as the categorical variable that is $0$ if $q(T_a)<1$ and $1$ if $q(T_a)>1$. Finally we have performed a logistic regression of $Y(T_a)$ against $\langle ES \rangle (T_a)$: namely,
we assume that \cite{statistical_learning_book}:

\begin{equation}\label{eq:logistic_regression}
 P\big\{Y(T_a)=1| \langle ES \rangle (T_a)=x\big\} = S\big(\beta_0+\beta_1 x \big) ~~,
\end{equation}
where $S(t)$ is the sigmoid function $S(t)=\frac{1}{1+e^{-t}}$ \cite{logistic_regression_book}; we estimate parameters $\beta_0$ and $\beta_1$ from the observations $\{ \langle ES \rangle (T_a) , q(T_a) \}_{a=1,...,n}$ through Maximum Likelihood \cite{machine_learning_book}. 

Once the model has been calibrated, given a new observation $\langle ES \rangle (T_{n+1}) = x$ we have predicted $Y(T_{n+1})=1$ if $P\big\{Y(T_{n+1})=1| \langle ES \rangle (T_{n+1}) = x\big\} > 0.5$, and $Y(T_{n+1})=0$ otherwise. This classification criterion, in a case with only one predictor, corresponds to classify $Y(T_{n+1})$ according to whether $\langle ES \rangle (T_{n+1})$ is greater or less than a threshold $r$ which depends on $\beta_0$ and $\beta_1$, as shown in Fig. \ref{fig:training_test} (right graphs) for a particular choice of parameters. Therefore the problem of predicting whether market volatility will increase or decrease boils down to a classification problem \cite{machine_learning_book} with $\langle ES \rangle (T_a)$ as predictor and $Y(T_a)$ as target variable. 

We have made use of a logistic regression because it is more suitable than a polynomial model for dealing with classification problems \cite{statistical_learning_book}. Other classification algorithms are available; we have chosen the logistic regression due to its simplicity. We have also implemented the KNN algorithm \cite{machine_learning_book} and we have found that it provides similar outcomes but worse results in terms of the forecasting performance metrics that we discuss in this section. 

We have then evaluated the goodness of the logistic regression at estimating $Y(T_{n+1})$ given a new observation $\langle ES \rangle (T_{n+1})$. 
To this end, we have computed three standard metrics for assessing the performance of a classification method: the probability of successful forecasting $P^{+}$, the True Positive Rate $TPR$ and the False Positive Rate $FPR$. $P^{+}$ represents the expected fraction of correct predictions, $TPR$ is the method goodness at identifying true positives (in this case, actual increases in volatility) and $FPR$ quantifies the method tendency to false positives (predictions of volatility increase when the volatility will actually decrease): see Methods for more details.  
Overall these metrics provide a complete summary of the model goodness at predicting changes in the market volatility \cite{statistical_learning_book}.   

In order to avoid overfitting we have estimated the metrics above by means of an out-of-sample procedure \cite{statistical_learning_book,machine_learning_book}. We have divided our dataset into two periods, a training set and a test set. In the training set we have calibrated the logistic equation in Eq. \ref{eq:logistic_regression}, estimating the parameters $\beta_0$ and $\beta_1$; in the test set we have used the calibrated model to measure the goodness of the model predictions by computing the measures of performance in Eq. \ref{eq:prob_forecast}-\ref{eq:fpr}. In Fig. \ref{fig:training_test} this division is shown for a particular choice of $\theta$ and $L$, for both NYSE and LSE dataset. In this example the percentage of data included in the test set (let us call it $f_{test}$) is $30\%$. 

Probabilities of successful forecasting $P^{+}$ are reported in Tabs. \ref{tab:nyse_p_outS} and \ref{tab:lse_p_outS}, for $f_{test}=30\%$.
As we can see $P^{+}$ is higher than $50\%$ for all combinations of parameters in NYSE dataset, and in almost all combinations for LSE dataset. 
Stars mark those values of $P^{+}$ that are significantly higher than the same probability obtained by using the most recent value of $q$ instead of $\langle ES \rangle (T_a)$ as a predictor for $q(T_a)$ in the logistic regression (let us call $P^{+}_q$ such probability). Specifically, we have defined a null model where variations from such probability $P^{+}_q$ are due to random fluctuations only; given $n$ observations, such fluctuations follow a Binomial distribution $B(P^{+}_q,n)$, with mean $n P^{+}_q$ and variance $n P^{+}_q(1-P^{+}_q)$. Then p-values have been calculated by using this null distribution for each combination of parameters. 
 This null hypothesis accounts for the predictability of $q(T_a)$ that is due to the autocorrelation of $q(T_a)$ only; therefore $P^{+}$ significantly higher than the value expected under this hypothesis implies a forecasting power of $\langle ES \rangle (T_a)$ that is not explained by the autocorrelation of $q(T_a)$. From the table we can see that $P^{+}$ is significant in 12 out of 16 combinations of parameters for NYSE dataset, and in 13 out of 16 for LSE dataset. This means that correlation persistence is a valuable predictor for future average correlation, able to outperform forecasting method based on past average correlation trends.
These results are robust against changes of $f_{test}$, as long as the training set is large enough to allow an accurate calibration of the logistic regression. We have found this condition is satisfied for $f_{test} < 40\%$.

However $P^{+}$ does not give any information on the method ability to distinguish between true and false positives. To investigate this aspect we need $TPR$ and $FPR$. A traditional way of representing both measures from a binary classifier is the so-called ``Receiver operating characteristic'' (ROC) curve \cite{roc_curve}. In a ROC plot, $TPR$ is plotted against $FPR$ as the discriminant threshold is varied. The discriminant threshold $p_{max}$ is the value of the probability in Eq. \ref{eq:logistic_regression} over which we classify $Y(T_a)=1$: the higher $p_{max}$ is, the less likely the method is to classify $Y(T_a)=1$ (in the analysis on $P^{+}$ we chose $p_{max}=0.5$). Ideally, a perfect classifier would yield $TPR=1$ for all $p_{max}>0$, whereas a random classifier is expected to lie on the line $TPR=FPR$. Therefore a ROC curve which lies above the line $TPR=FPR$ indicates a classifier that is better than chance at distinguishing true from false positives \cite{statistical_learning_book}. 

\begin{figure}[t!]
\hspace{-4em}
    \includegraphics[scale=0.4]{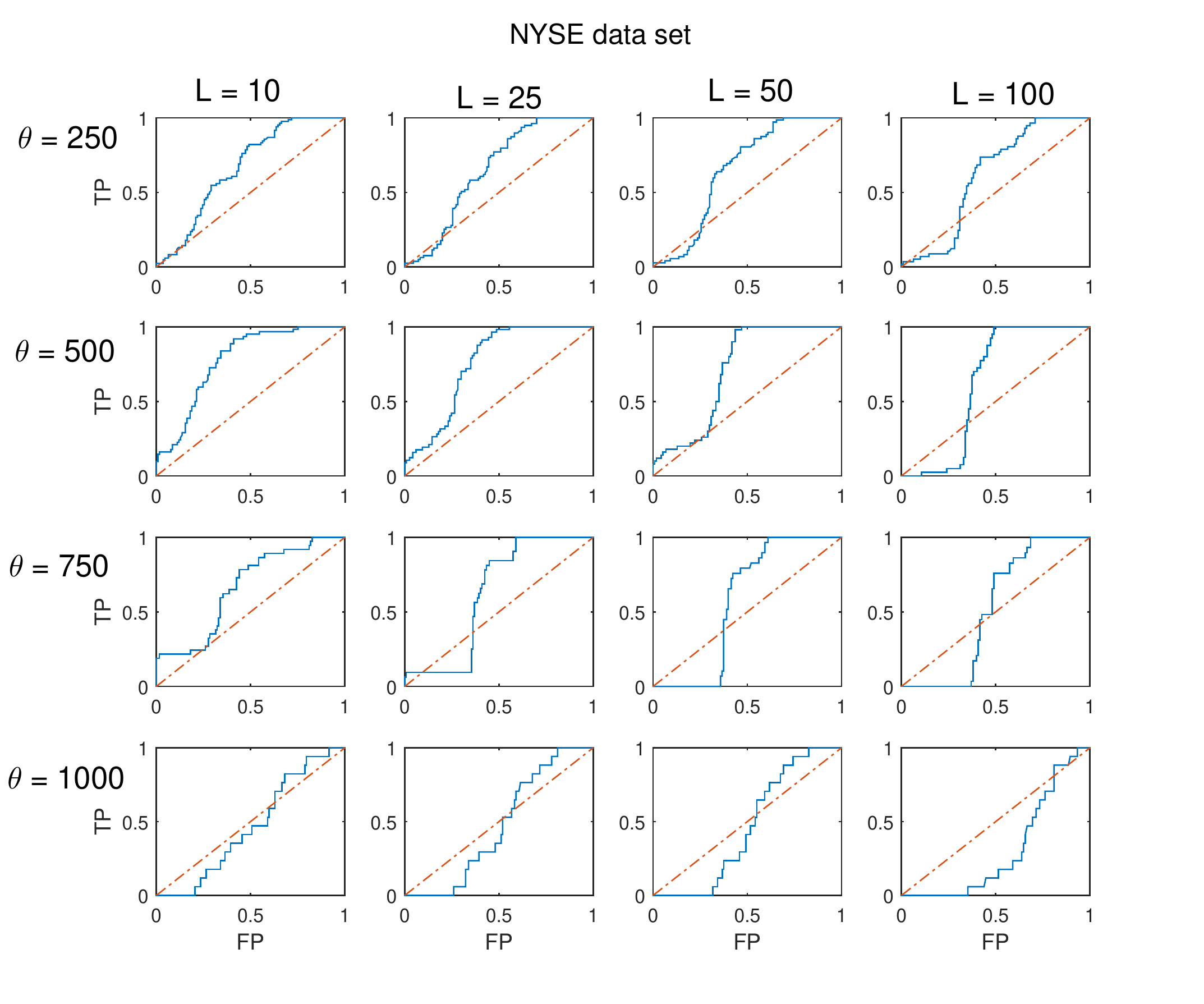}
    \includegraphics[scale=0.4]{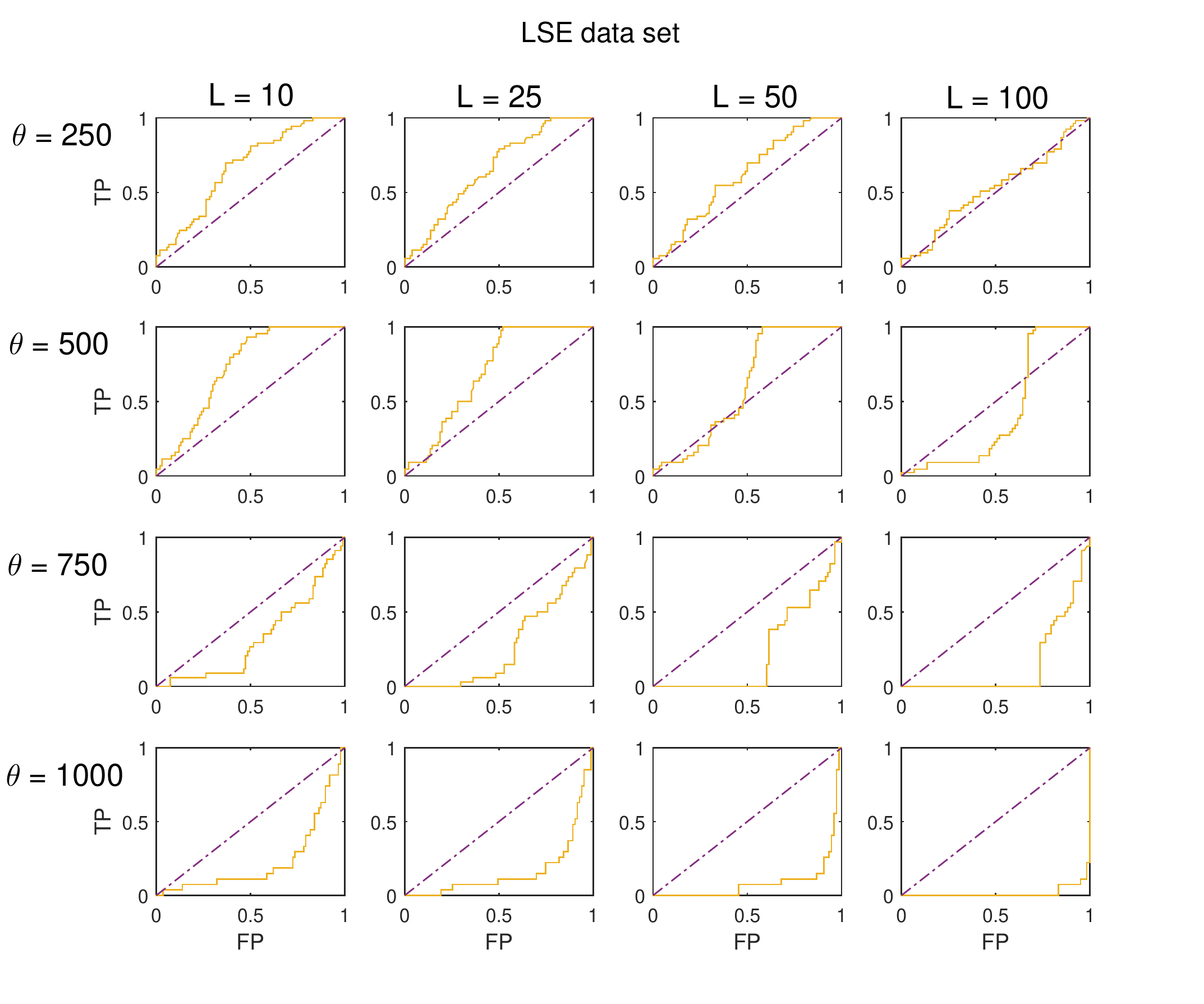}
\caption{ \label{fig:roc_curve} {\bf Receiver operating characteristic (ROC) curve}. Upper graph: True positive rate (TPR) against False positive rate (FPR) as the discriminant threshold $p_{max}$ of the classifier is varied, for each combination of parameters $\theta$ and $L$ in the NYSE dataset. The closer the curve is to the upper left corner of each graph, the better is the classifier compared to chance. Bottom graph: True positive rate (TPR) against False positive rate (FPR) as the discriminant threshold $p_{max}$ of the classifier is varied, for each combination of parameters $\theta$ and $L$ in the LSE dataset.}
\end{figure}

As one can see from Fig. \ref{fig:roc_curve}, the ROC curve's position depends on the choice of parameters $\theta$ and $L$. In this respect our classifier performs better for low values of $L$ and $\theta$. This can be quantified by measuring the area under the ROC curve; such measure, often denoted by AUC \cite{statistical_learning_book}, is shown in Tabs. \ref{tab:auc_nyse_outS}-\ref{tab:auc_lse_outS}. For both datasets the optimal choice of parameters is $\theta=500$ and $L=10$. 

\begin{table}[h!]\centering
\caption{\label{tab:nyse_p_outS} NYSE dataset: {\bf Probability of successful forecasting $P^{+}$}, for different combinations of parameters $\theta$ and $L$.
Out-of-sample analysis.}
\begin{tabular}{ cc | c c c c |}
\cline{3-6}
& & \multicolumn{4}{c|}{L} \\
\cline{3-6}
 & & \textbf{10} & \textbf{25} & \textbf{50} & \textbf{100}\\
\hline 
\multicolumn{1}{|c}{\multirow{4}{*}{$\theta$}} & \multicolumn{1}{|c|}{\textbf{250}}  &  0.546 &	0.560*	& 0.599** &	0.539**\\
 \multicolumn{1}{ |c  }{}   & \multicolumn{1}{|c|}{\textbf{500}}  &  0.704** &	0.695** &	0.658** &	0.605**\\
 \multicolumn{1}{ |c  }{} & \multicolumn{1}{|c|}{\textbf{750}}  & 0.634*	& 0.585 &	0.539 &	0.708*\\
 \multicolumn{1}{ |c  }{} & \multicolumn{1}{|c|}{\textbf{1000}} & 0.704* &	0.7638** &	0.839** &	0.860\\
\hline
\addlinespace[1ex]
\multicolumn{6}{l}{\textsuperscript{**}$p<0.001$, \textsuperscript{*}$p<0.01$,}
\end{tabular}
\end{table}

\begin{table}[h!]\centering
\caption{\label{tab:lse_p_outS} LSE dataset: {\bf Probability of successful forecasting $P^{+}$}, for different combinations of parameters $\theta$ and $L$.
Out-of-sample analysis.}
\begin{tabular}{ cc | c c c c |}
\cline{3-6}
& & \multicolumn{4}{c|}{L} \\
\cline{3-6}
 & & \textbf{10} & \textbf{25} & \textbf{50} & \textbf{100}\\
\hline 
\multicolumn{1}{|c}{\multirow{4}{*}{$\theta$}} & \multicolumn{1}{|c|}{\textbf{250}}  &  0.616**	& 0.645** &	0.612** &	0.568**\\
 \multicolumn{1}{ |c  }{}   & \multicolumn{1}{|c|}{\textbf{500}}  &  0.652** &	0.635** &	0.598** &	0.393\\
 \multicolumn{1}{ |c  }{} & \multicolumn{1}{|c|}{\textbf{750}}  & 0.651** &	0.560** &	0.453**	& 0.412\\
 \multicolumn{1}{ |c  }{} & \multicolumn{1}{|c|}{\textbf{1000}} & 0.544** &	0.573** &	0.706** &	0.689\\
\hline
\addlinespace[1ex]
\multicolumn{6}{l}{\textsuperscript{**}$p<0.001$, \textsuperscript{*}$p<0.01$,}
\end{tabular}
\end{table}

\begin{table}[t]\centering
\caption{\label{tab:auc_nyse_outS} NYSE dataset: {\bf Area under the curve (AUC)}, measured from the ROC curve in Fig. \ref{fig:roc_curve}. Values greater than 0.5 indicate that the classifier performs better than chance.}
\begin{tabular}{ cc | c c c c |}
\cline{3-6}
& & \multicolumn{4}{c|}{L} \\
\cline{3-6}
 & & \textbf{10} & \textbf{25} & \textbf{50} & \textbf{100}\\
\hline 
\multicolumn{1}{|c}{\multirow{4}{*}{$\theta$}} & \multicolumn{1}{|c|}{\textbf{250}}  &  0.669 &	0.652 &	0.655 &	0.616 \\
 \multicolumn{1}{ |c  }{}   & \multicolumn{1}{|c|}{\textbf{500}}  & 0.775 &	0.753 &	0.710 &	0.625\\
 \multicolumn{1}{ |c  }{} & \multicolumn{1}{|c|}{\textbf{750}}  & 0.663 &	0.6220 &	0.574 &	0.520\\
 \multicolumn{1}{ |c  }{} & \multicolumn{1}{|c|}{\textbf{1000}} & 0.467 &	0.470 &	0.462	& 0.314\\
\hline
\end{tabular}
\end{table} 

\begin{table}[t]\centering
\caption{\label{tab:auc_lse_outS} LSE dataset: {\bf Area under the curve (AUC)}, measured from the ROC curve in Fig. \ref{fig:roc_curve}. Values greater than 0.5 indicate that the classifier performs better than chance.}
\begin{tabular}{ cc | c c c c |}
\cline{3-6}
& & \multicolumn{4}{c|}{L} \\
\cline{3-6}
 & & \textbf{10} & \textbf{25} & \textbf{50} & \textbf{100}\\
\hline 
\multicolumn{1}{|c}{\multirow{4}{*}{$\theta$}} & \multicolumn{1}{|c|}{\textbf{250}}  &  0.673 &	0.658 &	0.618 &	0.524 \\
 \multicolumn{1}{ |c  }{}   & \multicolumn{1}{|c|}{\textbf{500}}  & 0.727 &	0.700 &	0.602 &	0.431\\
 \multicolumn{1}{ |c  }{} & \multicolumn{1}{|c|}{\textbf{750}}  & 0.324 &	0.274 &	0.234 &	0.148\\
 \multicolumn{1}{ |c  }{} & \multicolumn{1}{|c|}{\textbf{1000}} & 0.233 &	0.168 &	0.0918 &	0.0160\\
\hline
\end{tabular}
\end{table}

\subsection*{Temporal evolution of forecasting performance}

 \begin{figure}[ht!]
    \includegraphics[scale=0.4]{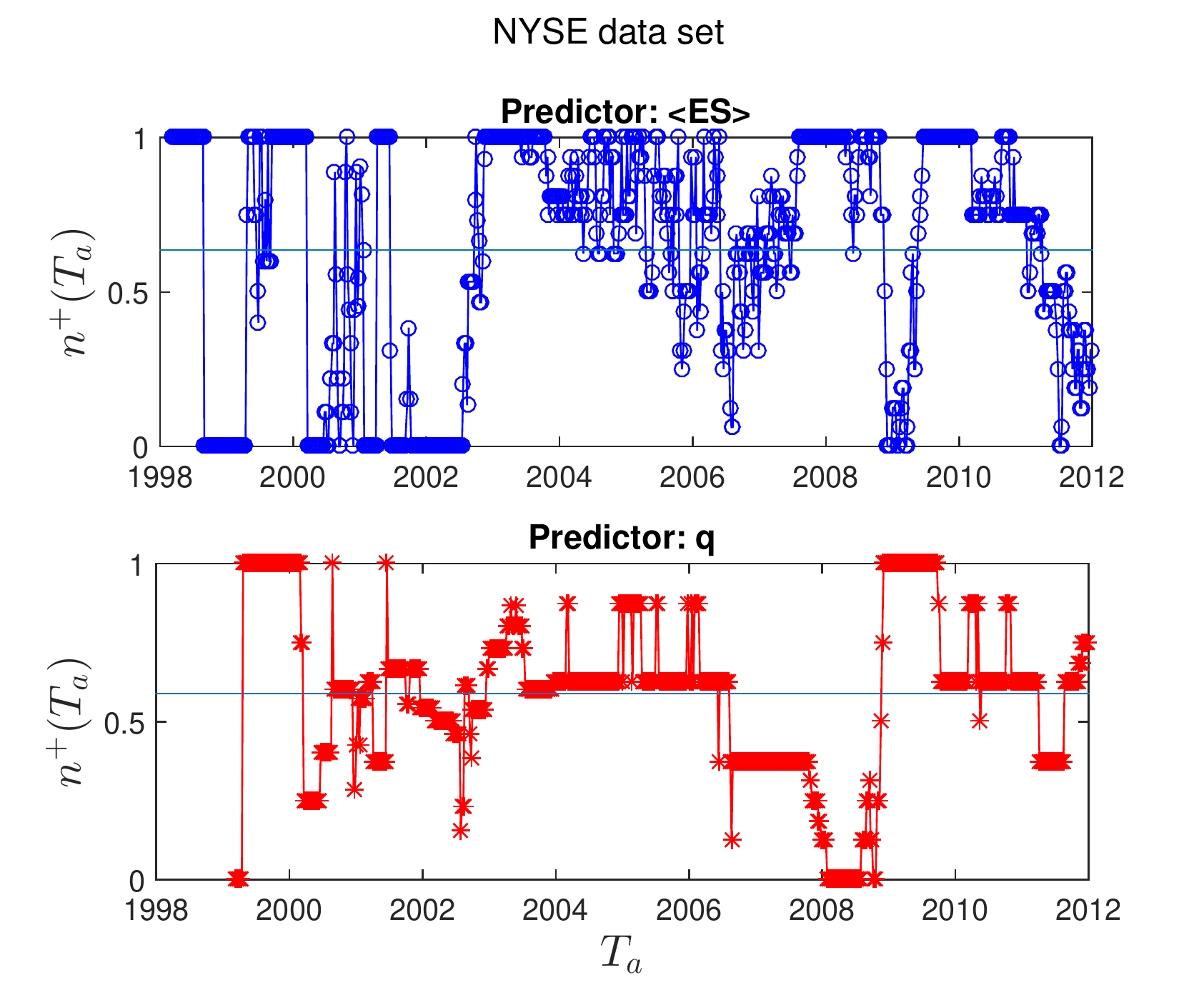}
    \includegraphics[scale=0.4]{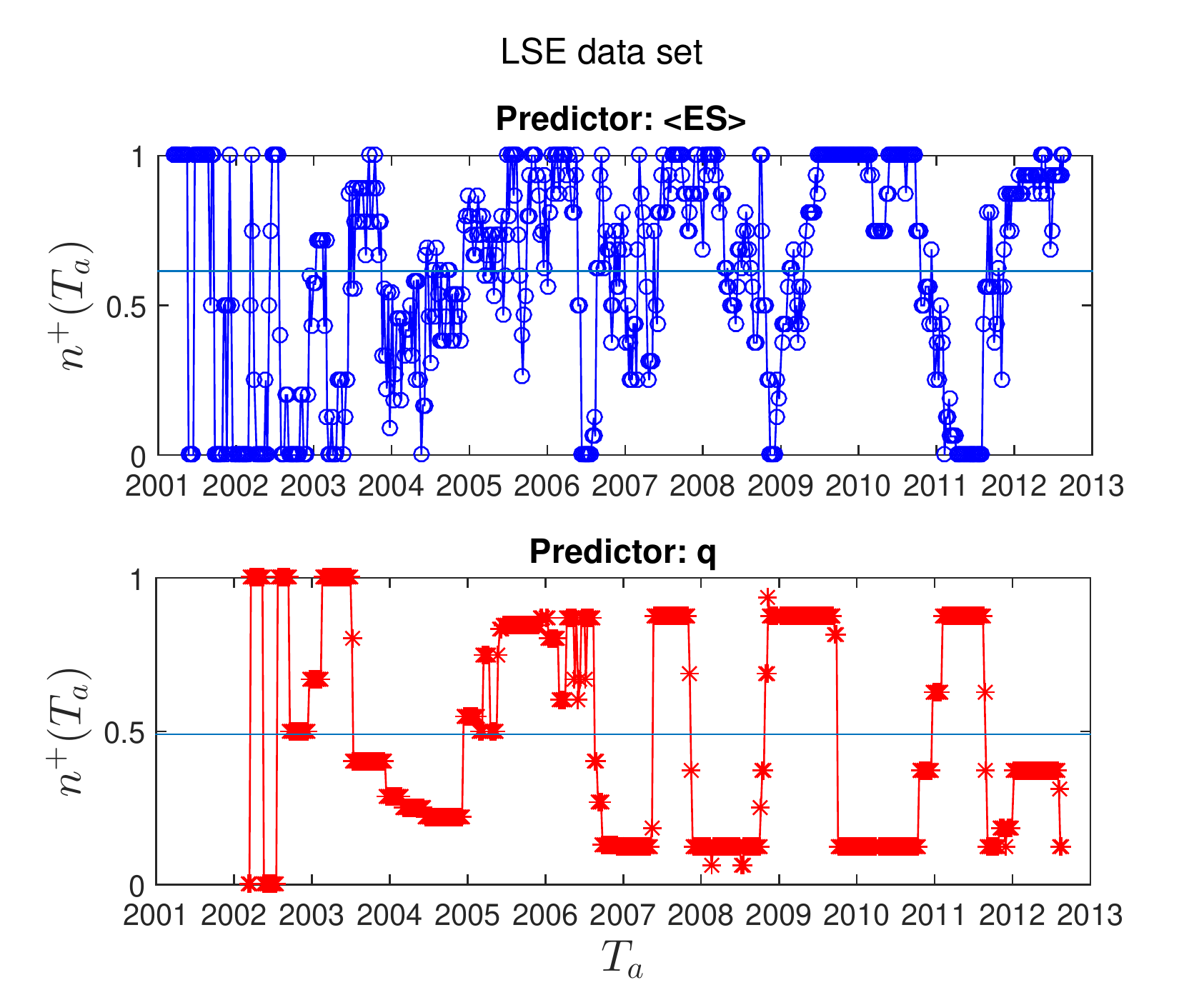}
	
\caption{\label{fig:forecast_time} {\bf Fraction of successful predictions as a function of time}. NYSE (left graph) and LSE dataset (right graph). Forecasting is based on logistic regression with predictor $\langle ES(T_a)\rangle$ (top graphs) and most recent value of $q(T_a)$ (bottom graphs). Horizontal lines represent the average over the entire period.}
\end{figure}			

In this section we look at how the forecasting performance changes at different time periods. In order to explore this aspect we have counted at each time window $T_a$ the number $N^{+}(T_a)$ of $Y(T_a)$ predictions (out of the 16 predictions corresponding to as many combinations of $\theta$ and $L$) that have turned out to be correct; we have then calculated the fraction of successful predictions $n^{+}(T_a)$ as $n^{+}(T_a)=N^{+}(T_a)/16$. In this way $n^{+}(T_a)$ is a proxy for the goodness of our method at each time window. Logistic regression parameters $\beta_0$ and $\beta_1$ have been calibrated by using the entire time period as training set, therefore this amounts to an in-sample analysis. 

In Fig. \ref{fig:forecast_time} we show the fraction of successful predictions for both NYSE and LSE datasets (upper graphs, blue circles). For comparison we also show the same measure obtained by using the most recent value of $q(T_a)$ as predictor (bottom graphs); as in the previous section, it represents a null model that makes prediction by using only the past evolution of $q(T_a)$. As we can see, both predictions based on $\langle ES \rangle (T_a)$ and on past values of $q(T_a)$ display performances changing in time. In particular $n^{+}(T_a)$ drops just ahead of the main financial crises (the market downturn in March 2002, 2007-2008 financial crisis, Euro zone crisis in 2011); this is probably due to the abrupt increase in volatility that occurred during these events and that the models took time to detect. After these drops though performances based on $\langle ES \rangle (T_a)$ recover much more rapidly than those based on past value of $q(T_a)$. For instance in the first months of 2007 our method shows quite high $n^{+}(T_a)$ (more than $60\%$ of successful predictions), being able to predict the sharp increase in volatility to come in 2008 while predictions based on $q(T_a)$ fail systematically until 2009. Overall, predictions based on correlation structure persistence appear to be more reliable (as shown by the average $n^{+}(T_a)$ over all time windows, the horizontal lines in the plot) and faster at detecting changes in market volatility.   

\section*{Discussion}
In this paper we have proposed a new tool for forecasting market volatility based on correlation-based information filtering networks and logistic regression, useful for risk and portfolio management. The advantage of our approach over traditional econometrics tools, such as multivariate GARCH and stochastic covariance models, is the ``top-down'' methodology that treats correlation matrices as the fundamental objects, allowing to deal with many assets simultaneously; in this way the curse of dimensionality, that prevents e.g. multivariate GARCH to deal with more than few assets, is overcome. We have proven the forecasting power of this tool by means of out-of-sample analyses on two different stock markets; the forecasting performance has been proven to be statistically significant against a null model, outperforming predictions based on past market correlation trends. Moreover we have measured the ROC curve and identified an optimal region of the parameters in terms of True Positive and False Positive trade-off. The temporal analysis indicates that our method is able to adapt to abrupt changes in the market, such as financial crises, more rapidly than methods based on past volatility.   

This forecasting tool relies on an empirical fact that we have reported in this paper for the first time. Specifically, we have shown that there is a deep interplay between market volatility and the rate of change of the correlation structure. The statistical significance of this relation has been assessed by means of a block-bootstrapping technique. An analysis based on metacorrelation has revealed that this interplay is better highlighted when filtering based on Planar Maximally Filtered Graphs is used to estimate the correlation structure persistence. 

This finding sheds new light into the dynamic of correlation. The topology of Planar Maximally Filtered Graphs depends on the ranking of the $N(N-1)/2$ pairs of cross-correlations; therefore an increase in the rate of change in PMFGs topology points out a faster change of this ranking. Our result indicates that such increase is typically followed by a rise in the market volatility, whereas decreases are followed by drops.
A possible interpretation of this is related to the dynamics of risk factors in the market. Indeed higher volatility in the market is associated to the emergence of a (possibly new) risk factor that makes the whole system unstable; such transition could be anticipated by a quicker change of the correlation ranking, triggered by the still emerging factor and revealed by the correlation structure persistence. Such persistence can therefore be a powerful tool for monitoring the emergence of new risks, valuable for a wide range of applications, from portfolio management to systemic risk regulation. Moreover this interpretation would open interesting connections with those
approaches to systemic risk that make use of Principal Component Analysis, monitoring the emergence of new risk factors by means of spectral methods \cite{pca_systemic_risk,pca_systemic_risk2}. We plan to investigate all these aspects in a future work.

\section*{Methods}
\subsection*{Metacorrelation as a measure of correlation structure persistence}
Given two correlation matrices $\{\rho_{ij}(T_a)\}$ and $\{\rho_{ij}(T_b)\}$ at two different time windows $T_a$ and $T_b$, their metacorrelation $z(T_a,T_b)$ is defined as follows: 

 \begin{equation}
  z(T_a,T_b) = \frac{\langle \rho_{ij}(T_a) \rho_{ij}(T_b) \rangle_{ij}}{\sqrt{[\langle \rho^2_{ij}(T_a) \rangle_{ij} - \langle \rho_{ij}(T_a) \rangle_{ij}^2][\langle \rho^2_{ij}(T_b) \rangle_{ij} - \langle \rho_{ij}(T_b) \rangle_{ij}^2] }} ,
 \label{eq:z}
 \end{equation}
where $\langle ... \rangle_{ij}$ is the average over all couples of stocks $i,j$.
Similarly to Eq. \ref{eq:es} we have then defined $z(T_a)$ as the weighted average over $L$ past time windows:

\begin{equation}
 \langle z \rangle (T_a) = \sum_{b = a -L }^{a - 1} \omega(T_b) z(T_a, T_b) .
\end{equation}

\subsection*{Measures of classification performance}
With reference to Figs. \ref{fig:training_test} b) and d), let us define the number of observations in each quadrant $Q_i$ ($i=1,2,3,4$) as $|Q_i|$. In the terminology of classification techniques \cite{machine_learning_book}, $|Q_1|$ is the number of True Positive (observations for which the model correctly predicted $Y(T_{a})=1$), $|Q_3|$ is the number of True Negative (observations for which the model correctly predicted $Y(T_{a})=0$), $|Q_2|$ the number of False Negative (observations for which the model incorrectly predicted $Y(T_{a})=0$) and $|Q_4|$ the number of False Positive (observations for which the model incorrectly predicted $Y(T_a)=1$). We have then computed the following measures of quality of classification, that are the standard metrics for assessing the performances of a classification method \cite{machine_learning_book}:

\begin{itemize}

\item {\bf Probability of successful forecasting ($P^{+}$)} \cite{machine_learning_book}: represents the method probability of a correct prediction, expressed as fraction of observed $\langle ES \rangle (T_a)$ values through which the method has successfully identified the correspondent value of $Y(T_{a})$. In classification problems, sometimes, the error rate $I$ is used \cite{statistical_learning_book}, which is simply $I=1-P^{+}$. $P^{+}$ is computed as follows: 

 \begin{equation}\label{eq:prob_forecast}
  P^{+} = \frac{|Q_1|+|Q_3|}{|Q_1|+|Q_2|+|Q_3|+|Q_4|} .
 \end{equation}

\item {\bf True Positive Rate ($TPR$)} \cite{machine_learning_book}: it is the probability of predicting $Y(T_a)=1$, conditional to the fact that the real $Y(T_a)$ is indeed $1$ (that is, to predict an increase in volatility when the volatility will indeed increase); it represents the method sensitivity to increase in volatility. It is also called ``recall'' \cite{statistical_learning_book}. In formula:

 \begin{equation}\label{eq:tpr}
  TPR = \frac{|Q_1|}{|Q_1|+|Q_2|} .
 \end{equation}

\item {\bf False Positive Rate ($FPR$)} \cite{machine_learning_book}: it is the probability of predicting $Y(T_a)=1$, conditional to the fact that the real $Y(T_a)$ is instead $0$ (that is, to predict an increase in volatility when the volatility will actually decrease). It is also called ``1-specificity'' \cite{statistical_learning_book}. In formula:

 \begin{equation}\label{eq:fpr}
  FPR = \frac{|Q_4|}{|Q_3|+|Q_4|} .
 \end{equation}

\end{itemize}


\begin{thebibliography}{10}
\expandafter\ifx\csname url\endcsname\relax
  \def\url#1{\texttt{#1}}\fi
\expandafter\ifx\csname urlprefix\endcsname\relax\def\urlprefix{URL }\fi
\providecommand{\bibinfo}[2]{#2}
\providecommand{\eprint}[2][]{\url{#2}}

\bibitem{risk_forecasting_book}
\bibinfo{author}{Daníelsson, J.}
\newblock \emph{\bibinfo{title}{Financial risk forecasting}}
  (\bibinfo{publisher}{Wiley}, \bibinfo{year}{2011}).

\bibitem{multivariate_models}
\bibinfo{author}{Hafner, C.~M.} \& \bibinfo{author}{Manner, H.}
\newblock \bibinfo{title}{Multivariate time series models for asset prices}.
\newblock \emph{\bibinfo{journal}{Handbook of Computational Finance}}
  \bibinfo{pages}{89--115} (\bibinfo{year}{2012}).

\bibitem{bouchaud_book}
\bibinfo{author}{Bouchaud, J.-P.} \& \bibinfo{author}{Potters, M.}
\newblock \emph{\bibinfo{title}{Theory of Financial Risk and Derivative
  Pricing: From Statistical Physics to Risk Management}}
  (\bibinfo{publisher}{Cambridge}, \bibinfo{year}{2000}).

\bibitem{preis_stress}
\bibinfo{author}{Preis, T.}, \bibinfo{author}{Kenett, D.~Y.},
  \bibinfo{author}{Stanley, H.~E.}, \bibinfo{author}{Helbing, D.} \&
  \bibinfo{author}{Ben-Jacob, E.}
\newblock \bibinfo{title}{Quantifying the behavior of stock correlations under
  market stress}.
\newblock \emph{\bibinfo{journal}{Scientific Reports}}
  \textbf{\bibinfo{volume}{2}}, \bibinfo{pages}{752} (\bibinfo{year}{2012}).

\bibitem{multigarch_survey}
\bibinfo{author}{Bauwens, L.}, \bibinfo{author}{Laurent, S.} \&
  \bibinfo{author}{Rombouts, J.}
\newblock \bibinfo{title}{Multivariate {GARCH} models: a survey}.
\newblock \emph{\bibinfo{journal}{Journal of Applied Econometrics}}
  \textbf{\bibinfo{volume}{21}}, \bibinfo{pages}{79--109}
  (\bibinfo{year}{2006}).

\bibitem{stochastic_volatility}
\bibinfo{author}{Clark, P.}
\newblock \bibinfo{title}{A subordinate stochastic process model with finite
  variance for speculative prices}.
\newblock \emph{\bibinfo{journal}{Econometrica}} \textbf{\bibinfo{volume}{41}},
  \bibinfo{pages}{135–155} (\bibinfo{year}{1973}).

\bibitem{realized_volatility}
\bibinfo{author}{Andersen, T.}, \bibinfo{author}{Bollerslev, T.},
  \bibinfo{author}{Diebold, F.} \& \bibinfo{author}{Labys, P.}
\newblock \bibinfo{title}{Forecasting realized volatility}.
\newblock \emph{\bibinfo{journal}{Econometrica}} \textbf{\bibinfo{volume}{71
  (2)}}, \bibinfo{pages}{579--625} (\bibinfo{year}{2003}).

\bibitem{mantegna1}
\bibinfo{author}{Mantegna, R.~N.}
\newblock \bibinfo{title}{Hierarchical structure in financial markets}.
\newblock \emph{\bibinfo{journal}{Eur. Phys. J. B}}
  \textbf{\bibinfo{volume}{11}}, \bibinfo{pages}{193} (\bibinfo{year}{1999}).

\bibitem{Tumminello05}
\bibinfo{author}{Tumminello, M.}, \bibinfo{author}{Aste, T.},
  \bibinfo{author}{Di{~}Matteo, T.} \& \bibinfo{author}{Mantegna, R.}
\newblock \bibinfo{title}{A tool for filtering information in complex systems}.
\newblock \emph{\bibinfo{journal}{Proc. Natl. Acad. Sci.}}
  \textbf{\bibinfo{volume}{102}}, \bibinfo{pages}{10421--10426}
  (\bibinfo{year}{2005}).

\bibitem{dynamic_networks_correlation}
\bibinfo{author}{Aste, T.} \& \bibinfo{author}{Matteo, T.~D.}
\newblock \bibinfo{title}{Dynamical networks from correlations}.
\newblock \emph{\bibinfo{journal}{Physica A}} \textbf{\bibinfo{volume}{370}},
  \bibinfo{pages}{156–161} (\bibinfo{year}{2006}).

\bibitem{asset_graphs}
\bibinfo{author}{Onnela, J.~P.}, \bibinfo{author}{Chakraborti, A.},
  \bibinfo{author}{Kaski, K.}, \bibinfo{author}{Kert\'esz, J.} \&
  \bibinfo{author}{Kanto, A.}
\newblock \bibinfo{title}{Asset trees and asset graphs in financial markets}.
\newblock \emph{\bibinfo{journal}{Phys. Scr.}} \textbf{\bibinfo{volume}{T106}},
  \bibinfo{pages}{48} (\bibinfo{year}{2003}).

\bibitem{clustering_dyn_asset_graph}
\bibinfo{author}{Onnella, J.-P.}, \bibinfo{author}{Kaski, K.} \&
  \bibinfo{author}{Kert\'esz, J.}
\newblock \bibinfo{title}{Clustering and information in correlation based
  financial networks}.
\newblock \emph{\bibinfo{journal}{Eur. Phys. J. B}}
  \textbf{\bibinfo{volume}{38}}, \bibinfo{pages}{353--362}
  (\bibinfo{year}{2004}).

\bibitem{buccheri}
\bibinfo{author}{Buccheri, G.}, \bibinfo{author}{Marmi, S.} \&
  \bibinfo{author}{Mantegna, R.~N.}
\newblock \bibinfo{title}{Evolution of correlation structure of industrial
  indices of {U.S.} equity markets}.
\newblock \emph{\bibinfo{journal}{Phys. Rev. E}} \textbf{\bibinfo{volume}{88}},
  \bibinfo{pages}{012806} (\bibinfo{year}{2013}).

\bibitem{caldarelli_book}
\bibinfo{author}{Caldarelli, G.}
\newblock \emph{\bibinfo{title}{Scale-Free Networks}}
  (\bibinfo{publisher}{Oxford Finance Series}, \bibinfo{year}{2007}).

\bibitem{black_monday}
\bibinfo{author}{Onnela, J.~P.}, \bibinfo{author}{Chakraborti, A.},
  \bibinfo{author}{Kaski, K.} \& \bibinfo{author}{Kert\'esz, J.}
\newblock \bibinfo{title}{Dynamic asset trees and black monday}.
\newblock \emph{\bibinfo{journal}{Physica A}} \textbf{\bibinfo{volume}{324}},
  \bibinfo{pages}{247--252} (\bibinfo{year}{2003}).

\bibitem{cluster_portfolio}
\bibinfo{author}{Tola, V.}, \bibinfo{author}{Lillo, F.},
  \bibinfo{author}{Gallegati, M.} \& \bibinfo{author}{Mantegna, R.}
\newblock \bibinfo{title}{Cluster analysis for portfolio optimization}.
\newblock \emph{\bibinfo{journal}{J. Econ. Dyn. Control}}
  \textbf{\bibinfo{volume}{32}}, \bibinfo{pages}{235--258}
  (\bibinfo{year}{2008}).

\bibitem{bonanno_mst}
\bibinfo{author}{Bonanno, G.}, \bibinfo{author}{Caldarelli, G.},
  \bibinfo{author}{Lillo, F.} \& \bibinfo{author}{Mantegna, R.}
\newblock \bibinfo{title}{Topology of correlation-based minimal spanning trees
  in real and model markets}.
\newblock \emph{\bibinfo{journal}{Phys Rev E Stat Nonlin Soft Matter Phys}}
  \textbf{\bibinfo{volume}{68}}, \bibinfo{pages}{046130}
  (\bibinfo{year}{2003}).

\bibitem{invest_periph}
\bibinfo{author}{Pozzi, F.}, \bibinfo{author}{Di~Matteo, T.} \&
  \bibinfo{author}{Aste, T.}
\newblock \bibinfo{title}{Spread of risk across financial markets: better to
  invest in the peripheries}.
\newblock \emph{\bibinfo{journal}{Sci. Rep.}} \textbf{\bibinfo{volume}{3}},
  \bibinfo{pages}{1665} (\bibinfo{year}{2013}).

\bibitem{musmeci_DBHT}
\bibinfo{author}{Musmeci, N.}, \bibinfo{author}{Aste, T.} \&
  \bibinfo{author}{Di~Matteo, T.}
\newblock \bibinfo{title}{Relation between financial market structure and the
  real economy: Comparison between clustering methods}.
\newblock \emph{\bibinfo{journal}{PLoS ONE}} \textbf{\bibinfo{volume}{10 (4)}},
  \bibinfo{pages}{e0126998. doi: 10.1371/journal.pone.0126998}
  (\bibinfo{year}{2015}).

\bibitem{musmeci_jntf}
\bibinfo{author}{Musmeci, N.}, \bibinfo{author}{Aste, T.} \&
  \bibinfo{author}{Di{~}Matteo, T.}
\newblock \bibinfo{title}{Risk diversification: a study of persistence with a
  filtered correlation-network approach}.
\newblock \emph{\bibinfo{journal}{Journal of Network Theory in Finance}}
  \textbf{\bibinfo{volume}{1}}, \bibinfo{pages}{1--22} (\bibinfo{year}{2015}).

\bibitem{markowitz}
\bibinfo{author}{Markowitz, H.}
\newblock \bibinfo{title}{Portfolio selection}.
\newblock \emph{\bibinfo{journal}{The Journal of Finance}}
  \textbf{\bibinfo{volume}{7 (1)}}, \bibinfo{pages}{77--91}
  (\bibinfo{year}{1952}).

\bibitem{rmt_1}
\bibinfo{author}{Plerou, V.} \emph{et~al.}
\newblock \bibinfo{title}{Random matrix approach to cross-correlations in
  financial data}.
\newblock \emph{\bibinfo{journal}{Phys. Rev. E}} \textbf{\bibinfo{volume}{65}},
  \bibinfo{pages}{066126} (\bibinfo{year}{2002}).

\bibitem{pca_systemic_risk}
\bibinfo{author}{Kritzman, M.}, \bibinfo{author}{Li, Y.},
  \bibinfo{author}{Page, S.} \& \bibinfo{author}{Rigobon, R.}
\newblock \bibinfo{title}{Principal components as a measure of systemic risk}.
\newblock \emph{\bibinfo{journal}{Journal of Portfolio Management}}
  \textbf{\bibinfo{volume}{37}}, \bibinfo{pages}{112--126}
  (\bibinfo{year}{2011}).

\bibitem{pca_systemic_risk2}
\bibinfo{author}{Zheng, Z.}, \bibinfo{author}{Podobnik, B.},
  \bibinfo{author}{Feng, L.} \& \bibinfo{author}{Li, B.}
\newblock \bibinfo{title}{Changes in cross-correlations as an indicator for
  systemic risk}.
\newblock \emph{\bibinfo{journal}{Sci. Rep. 2}} \bibinfo{pages}{888;
  DOI:10.1038 srep00888} (\bibinfo{year}{2011}).

\bibitem{exp_smoothing}
\bibinfo{author}{Pozzi, F.}, \bibinfo{author}{Di~Matteo, T.} \&
  \bibinfo{author}{Aste, T.}
\newblock \bibinfo{title}{Exponential smoothing weighted correlations}.
\newblock \emph{\bibinfo{journal}{Eur. Phys. J. B}}
  \textbf{\bibinfo{volume}{85}}, \bibinfo{pages}{6} (\bibinfo{year}{2012}).

\bibitem{PMFG2}
\bibinfo{author}{Aste, T.}, \bibinfo{author}{Di~Matteo, T.} \&
  \bibinfo{author}{Hyde, S.~T.}
\newblock \bibinfo{title}{Complex networks on hyperbolic surfaces}.
\newblock \emph{\bibinfo{journal}{Physica A}} \textbf{\bibinfo{volume}{346}},
  \bibinfo{pages}{20} (\bibinfo{year}{2005}).

\bibitem{NJP10}
\bibinfo{author}{Aste, T.}, \bibinfo{author}{Shaw, W.} \&
  \bibinfo{author}{Di~Matteo, T.}
\newblock \bibinfo{title}{Correlation structure and dynamics in volatile
  markets}.
\newblock \emph{\bibinfo{journal}{New J. Phys.}} \textbf{\bibinfo{volume}{12}},
  \bibinfo{pages}{085009} (\bibinfo{year}{2010}).

\bibitem{kondor}
\bibinfo{author}{Pafka, S.} \& \bibinfo{author}{Kondor, I.}
\newblock \bibinfo{title}{Noisy covariance matrices and portfolio
  optimization}.
\newblock \emph{\bibinfo{journal}{Eur. Phys. J. B}}
  \textbf{\bibinfo{volume}{27}}, \bibinfo{pages}{DOI: 10.1140/epjb/e20020153,
  277--280} (\bibinfo{year}{2002}).

\bibitem{kondor2}
\bibinfo{author}{Pafka, S.} \& \bibinfo{author}{Kondor, I.}
\newblock \bibinfo{title}{Noisy covariance matrices and portfolio optimization
  {II}}.
\newblock \emph{\bibinfo{journal}{Physica A}} \textbf{\bibinfo{volume}{319}},
  \bibinfo{pages}{487 – 494} (\bibinfo{year}{2003}).

\bibitem{non_stationary_corr}
\bibinfo{author}{Livan, G.}, \bibinfo{author}{Inoue, J.} \&
  \bibinfo{author}{Scalas, E.}
\newblock \bibinfo{title}{On the non-stationarity of financial time series:
  impact on optimal portfolio selection}.
\newblock \emph{\bibinfo{journal}{J. Stat. Mech.}} \bibinfo{pages}{P07025}
  (\bibinfo{year}{2012}).

\bibitem{fisher_transform}
\bibinfo{author}{Fisher, R.}
\newblock \bibinfo{title}{On the `probable error' of a coefficient of
  correlation deduced from a small sample}.
\newblock \emph{\bibinfo{journal}{Metron}} \textbf{\bibinfo{volume}{1}},
  \bibinfo{pages}{3--32} (\bibinfo{year}{1921}).

\bibitem{math_statistics}
\bibinfo{author}{Kenney, J.~F.} \& \bibinfo{author}{Keeping, E.~S.}
\newblock \emph{\bibinfo{title}{Mathematics of Statistics}}
  (\bibinfo{publisher}{Princeton, NJ: Van Nostrand}, \bibinfo{year}{1947}).

\bibitem{block_bootstrapping}
\bibinfo{author}{Kunsch, H.}
\newblock \bibinfo{title}{The jackknife and the bootstrap for general
  stationary observations}.
\newblock \emph{\bibinfo{journal}{Ann. Statist.}}
  \textbf{\bibinfo{volume}{17}}, \bibinfo{pages}{1217–1241}
  (\bibinfo{year}{1989}).

\bibitem{bootstrapping}
\bibinfo{author}{Efron, B.}
\newblock \bibinfo{title}{Bootstrap methods: Another look at the jackknife}.
\newblock \emph{\bibinfo{journal}{Ann. Stat.}} \textbf{\bibinfo{volume}{7}},
  \bibinfo{pages}{1--26} (\bibinfo{year}{1979}).

\bibitem{optimal_block_bootstrapping}
\bibinfo{author}{Politis, D.~N.} \& \bibinfo{author}{White, H.}
\newblock \bibinfo{title}{Automatic block-length selection for the dependent
  bootstrap}.
\newblock \emph{\bibinfo{journal}{Econometrics Reviews}}
  \textbf{\bibinfo{volume}{23 (1)}}, \bibinfo{pages}{53–70}
  (\bibinfo{year}{2004}).

\bibitem{mst_history}
\bibinfo{author}{Graham, R.} \& \bibinfo{author}{Hell, P.}
\newblock \bibinfo{title}{On the history of the minimum spanning tree problem}.
\newblock \emph{\bibinfo{journal}{Annals of the History of Computing}}
  \textbf{\bibinfo{volume}{7 (1)}}, \bibinfo{pages}{43–57}
  (\bibinfo{year}{1985}).

\bibitem{statistical_learning_book}
\bibinfo{author}{James, G.}, \bibinfo{author}{Witten, D.},
  \bibinfo{author}{Hastie, T.} \& \bibinfo{author}{Tibshirani, R.}
\newblock \emph{\bibinfo{title}{An Introduction to Statistical Learning with
  Applications in R}} (\bibinfo{publisher}{Springer}, \bibinfo{year}{2014}).

\bibitem{logistic_regression_book}
\bibinfo{author}{Hilbe, J.~M.}
\newblock \emph{\bibinfo{title}{Logistic Regression Models}}
  (\bibinfo{publisher}{Chapman \& Hall/CRC Press}, \bibinfo{year}{2009}).

\bibitem{machine_learning_book}
\bibinfo{author}{Bishop, C.}
\newblock \emph{\bibinfo{title}{Pattern Recognition and Machine Learning}}
  (\bibinfo{publisher}{Springer}, \bibinfo{year}{2007}).

\bibitem{roc_curve}
\bibinfo{author}{Spackman, K.~A.}
\newblock \bibinfo{title}{Signal detection theory: Valuable tools for
  evaluating inductive learning}.
\newblock \emph{\bibinfo{journal}{Proceedings of the Sixth International
  Workshop on Machine Learning. San Mateo, CA: Morgan Kaufmann}}
  \bibinfo{pages}{160–163} (\bibinfo{year}{1989}).

\end{thebibliography}

%

\section*{Acknowledgements}
The authors wish to thank Bloomberg for providing the data.
TDM wishes to thank the COST Action TD1210 for partially supporting this work. TA acknowledges support of the UK Economic and Social Research Council (ESRC) in funding the Systemic Risk Centre (ES/K002309/1). 

\section*{Additional information}

\textbf{Competing financial interests.} The authors declare no
competing financial interest.

\end{document}